\documentclass[nofootinbib,aps,prd,groupedaddress,preprintnumbers,%
  showpacs,showkeys,floatfix,amssymb,amsfonts]{revtex4-1}  
\usepackage{hyperref}
\usepackage[usenames]{color}
\usepackage{todonotes}
\usepackage{bbold}
\usepackage{amsmath}
\usepackage{multirow}
\usepackage{graphicx}
\usepackage{xfrac}
\usepackage[utf8]{inputenc}
 
\usepackage{amsfonts}
\usepackage{amssymb}
\usepackage{extarrows}

\newcommand{\rr}{\langle r^2 \rangle}

\newcommand{\be}{\begin{equation}}
\newcommand{\ee}{\end{equation}}
\newcommand{\bea}{\begin{eqnarray}}
\newcommand{\eea}{\end{eqnarray}}

\newcommand{\Op}{{\cal O}}
\newcommand{\avgx}{\langle x \rangle}
\newcommand{\avgxx}{\langle x^2 \rangle}
\newcommand{\Dlr}{\buildrel \leftrightarrow \over D\raise-1pt\hbox{}}

\newcommand{\avgxxx}{\langle x^3 \rangle}

\definecolor{Teal}{HTML}{008080}
\newcommand{\alx}[1]{\textcolor{Teal}{#1}} 
\newcommand{\del}[1]{\textcolor{gray}{#1}} 

\begin{document}

\title{The scalar, vector and tensor form factors for the pion and kaon from lattice QCD}

\author{
\vspace*{0.35cm}
  Constantia Alexandrou$^{1,2}$,
  Simone Bacchio$^{2}$,
  Ian Clo\"et$^3$,
  Martha Constantinou$^{4}$,\\[1ex]
  Joseph Delmar$^{4}$,
Kyriakos Hadjiyiannakou$^{1,2}$,
  Giannis Koutsou$^{2}$,
  Colin Lauer$^4$,
  Alejandro Vaquero$^5$ \\[2ex]
 (ETM Collaboration)
}

\affiliation{
  \vskip 0.25cm
  $^1$Department of Physics, University of Cyprus,  P.O. Box 20537,  1678 Nicosia, Cyprus\\
  \vskip 0.05cm
   $^2$Computation-based Science and Technology Research Center,
  The Cyprus Institute, 20 Kavafi Str., Nicosia 2121, Cyprus \\
 \vskip 0.05cm
  $^3$ Physics Division, Argonne National Laboratory, Lemont, Illinois 60439, USA\\
  \vskip 0.05cm
  $^4$Department of Physics, Temple University, 1925 N. 12th Street, Philadelphia, PA 19122-1801, USA\\ 
  \vskip 0.05cm
  $^5$Department of Physics and Astronomy, University of Utah, Salt Lake City, Utah 84112, USA\\
  \vskip 0.05cm
 \phantom{-}
\centerline{\today}
 {\includegraphics[scale=0.2]{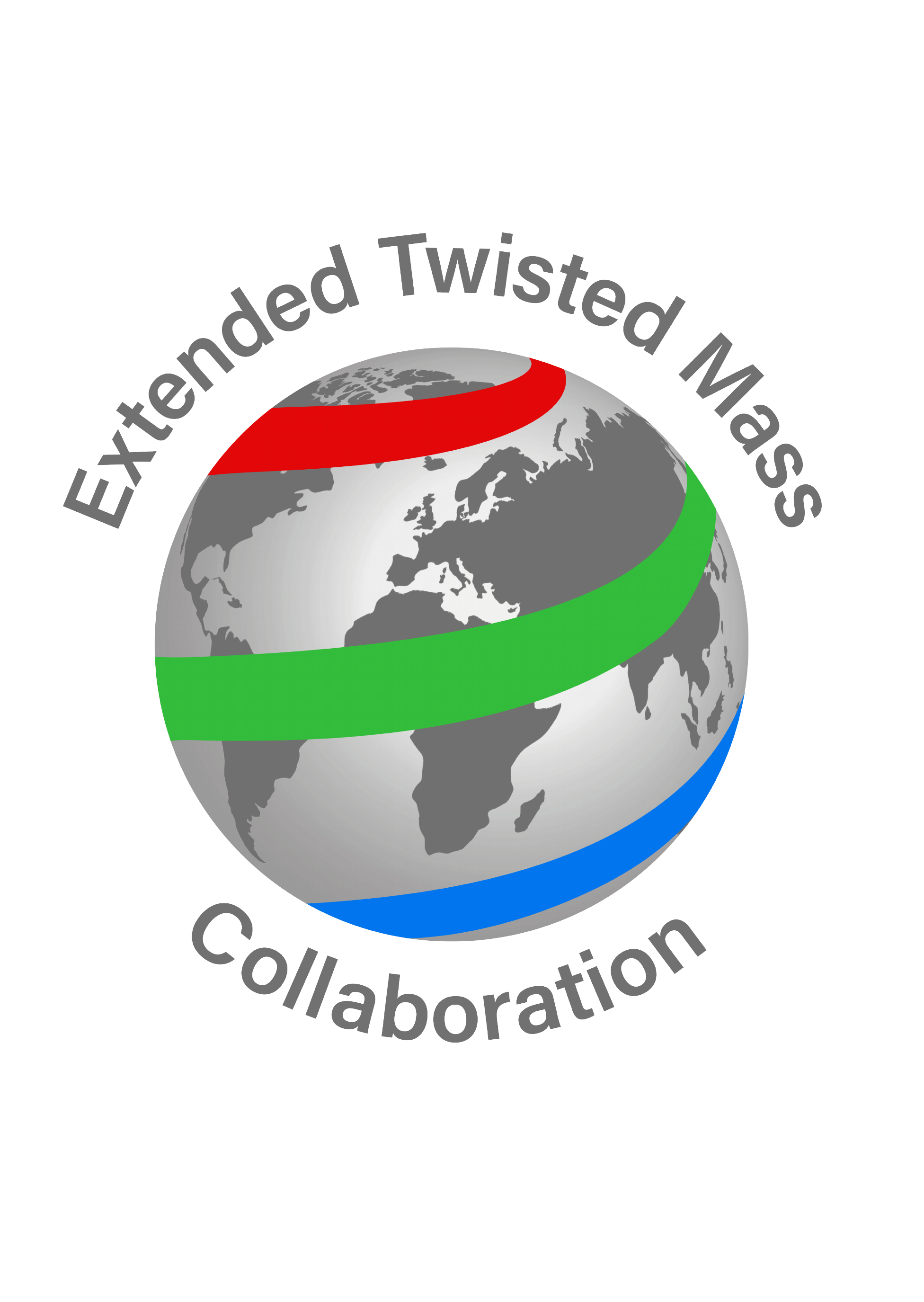}}
  }

\begin{abstract}
We present a calculation of the scalar, vector and tensor form factors for the pion and kaon in lattice QCD. We use an ensemble of two degenerate light, a strange and a charm quark ($N_f=2+1+1$) of maximally twisted mass fermions with clover improvement. The corresponding pion and kaon masses are about 265 MeV and 530 MeV, respectively. The calculation is done in both rest and boosted frames obtaining data for four-vector momentum transfer squared up to $-q^2=2.5$ GeV$^2$ for the pion and 3~GeV$^2$ for the kaon. The excited-states effects are studied by analyzing six values of the source-sink time separation for the rest frame ($1.12-2.23$~fm) and for four values for the boosted frame ($1.12-1.67$~fm). The lattice data are renormalized non-perturbatively and the results for the scheme- and scale-dependent scalar and tensor form factors are presented in the $\overline{\rm MS}$ scheme at a scale of 2 GeV. We apply different parametrizations to describe $q^2$-dependence of the form factors to extract the scalar, vector and tensor radii, as well as the  tensor anomalous magnetic moment. We compare the pion and kaon form factors to study SU(3) flavor symmetry breaking effects. By combining the data for the vector and tensor form factors we also obtain the lowest moment of the densities of transversely polarized quarks in the impact parameter space. Finally, we give an estimate for the average transverse shift in the $y$ direction for polarized quarks in the $x$ direction.
\end{abstract}

\maketitle

\section{Introduction}

Quantum chromodynamics (QCD) is the theory that best describes the strong interactions. It accommodates a complex array of strongly interacting phenomena that requires, at low energies, a non-perturbative framework. One of them is the emergent hadronic mass (EHM), a mechanism usually invoked to explain the large masses of  hadrons~\cite{Cui:2020dlm,Roberts:2021nhw}. Another particular property of QCD is the spontaneous chiral symmetry breaking (SCSB). In the absence of the Higgs interaction, the quarks would be massless, and the QCD Lagrangian would enjoy full chiral symmetry. But, we know this symmetry is spontaneously broken, giving rise to several  Goldstone bosons, i.e. the pions and the kaons. The lightest quarks have small enough masses to make chiral symmetry a good approximation in certain cases, and hence we expect the would-be Goldstone bosons to have small masses, smaller than what one would expect from the EHM mechanism, but still larger than the sum of the masses of their valence constituents. Therefore, there is an interplay between the EHM and the Higgs mechanisms, with the former decreasing the masses of the Goldstone bosons, while the latter increases their masses.
Exploring the structure of these light mesons can thus cast light into the EHM mechanism, and how it interacts with the Higgs sector. It is also interesting to study the pion and kaon structure and compare to the proton. For instance, even in the absence of the quark coupling to the Higgs boson, the proton has a non-zero mass, while the pion and kaon would be massless.

Furthermore, pions and kaons are very important for describing the long-range dynamics of the strong interactions. Studying their structure is of foremost importance to better understand QCD dynamics~\cite{Hagler:2009ni}. In fact, the importance of these mesons justifies the intense experimental activity since the 1970's for the pion~\cite{Dally:1977vt,Dally:1981ur,Dally:1982zk,Palestini:1985zc,Amendolia:1983di,Amendolia:1984nz,Amendolia:1986wj,Amendolia:1985bs,Ackermann:1977rp,Brauel:1979zk,Bebek:1977pe,Bebek:1974ww,Bebek:1974iz,Brown:1973wr,JeffersonLabFpi:2000nlc,JeffersonLabFpi:2007vir,JeffersonLabFpi-2:2006ysh,JeffersonLab:2008gyl,JeffersonLab:2008jve,Horn:2007ug,Belle:2008xpe,Barkov:1985ac,GoughEschrich:2001ji}. In addition,   several  studies exist using chiral perturbation theories and other phenomenological approaches ~\cite{Gasser:1990bv,Bijnens:1998fm,Guerrero:1997ku,Caprini:1999ws,Ananthanarayan:2012tn,Chang:2013nia,Dubnicka:2014xga,Hutauruk:2016sug,Ananthanarayan:2017efc,Hutauruk:2018zfk}, as well as, lattice calculations on the pion (vector) form factor (see, e.g., Refs.~\cite{Brommel:2006ww,Boyle:2008yd,JLQCD:2009ofg,Bali:2013gya,Fukaya:2014jka,Colangelo:2018mtw,Wang:2020nbf,Gao:2021xsm}).
Almost all information on  pion structure comes from using the electromagnetic current  as a probe, while the scalar and tensor form factors are lesser studied. Existing lattice calculations for the scalar form factor of the pion can be found in Refs. ~\cite{Kaneko:2008kx,JLQCD:2009ofg,Kaneko:2010ru,Gulpers:2013uca} and for the tensor in Ref.~\cite{Brommel:2007xd}. The work of Ref.~\cite{Kaneko:2010ru} includes a calculation of the kaon vector form factor.
The kaon has always been more elusive, as compared to the  pion, as the pion is more accessible in experiments and theoretical studies using chiral perturbation theory. Experiments involving kaons are harder, especially at large momentum. To date, only a few measurements exist~\cite{Dally:1980dj,Amendolia:1986ui,Brauel:1979zk,Carmignotto:2018uqj}. Nonetheless, the forthcoming JLAB E12-09-001 experiment and the Electron-Ion Collider (EIC) aim to generate a large amount of high precision data for the kaon~\cite{Arrington:2021biu}.

In this work, we calculate the scalar, vector and tensor form factors of both the pion and the kaon. We neglect a certain class of sea quark contributions connected to the so-called disconnected diagrams. These are expected to be small, especially for larger than physical pion mass. Of particular interest is the $q^2$-dependence of the form factors, which leads to the monopole masses and radii when a parametrization is applied. For the tensor case, we can also extract  the tensor anomalous magnetic moment. We combine the lattice data on the vector and tensor form factors, to extract information on the transverse spin of the mesons under study. We draw qualitative conclusions on the SU(3) flavor symmetry breaking effect, by comparing the form factors between the pion and kaon. 

The article is structured as follows. Section~\ref{sec_lat} discusses the theoretical approach and the lattice setup providing detailed information on the ensemble we use and the kinematical frames. Section~\ref{sec:renorm} summarizes the renormalization procedure and gives results on the renormalization factors. In Section~\ref{sec:analysis} we explain how we extract the matrix elements including details on the excited states analysis. Our results in the rest and boosted frames are provided in Sections~\ref{sec_pion} and \ref{sec_kaon} for the pion and the kaon form factors, respectively. These sections also include the parametrization of the $q^2=-Q^2$ dependence, results on the monopole masses, the tensor anomalous magnetic moment and the radii. Using the parametrizations of the pion and kaon form factors, we study SU(3) flavor symmetry breaking effects, which are presented in Section~\ref{sec:SU3}. Section~\ref{sec:impact_b} describes the framework for studying the transverse spin structure of the pion and the kaon and the extraction of the average transverse shift in the $y$ direction for polarized quarks in the $x$ direction. Finally, Section~\ref{sec:summary} summarizes our results and gathers our conclusions.

\section{Theoretical and Lattice Setup}
\label{sec_lat}

The form factors for a particular quark flavor $f$ are obtained from the matrix elements of ultra-local operators
\begin{equation}
    \langle M({p}') | \Op^f_\Gamma | M({p}) \rangle \,,
\end{equation}
where the allowed operator structure $\Op^f_\Gamma $ for 0-spin mesons are the scalar, $\Op^f_S=\bar{\psi}\hat{1}\psi$, vector, $\Op^f_V=\bar{\psi}\gamma^\mu\psi$, and tensor, $\Op^f_T=\bar{\psi}\sigma^{\mu\nu}\psi$ with  $\sigma^{\mu\nu}=\frac{1}{2}[\gamma^\mu,\gamma^\nu]$. The 4-vector momentum transfer ($t\equiv-Q^2$) dependence of the form factor is extracted from the off-forward matrix element, where the momentum transfer between the initial (${p}$) and final (${p'}$) state is ${q}={p'} - {p}$. The decomposition of each matrix element for the general frame in Euclidean space is~\cite{Hagler:2009ni}
\begin{align}
    \langle M({p}') | \Op^f_S | M({p}) \rangle &= \frac{1}{\sqrt{4 E(p) E(p')}} A^{M^f}_{S10}\,, \\[1ex]
    \langle M({p}') | \Op^f_{V^\mu} | M({p}) \rangle &= -i\, \frac{2\, P^\mu}{\sqrt{4 E(p) E(p')}} \, A^{M^f}_{10}\,, \\[1ex]
    \langle M({p}') | \Op^f_{T^{\mu\nu}} | M({p}) \rangle &= i\, \frac{(P^\mu q^\nu - P^\nu q^\mu)}{m_M \sqrt{4 E(p) E(p')}}  \,B^{M^f}_{T10}\,.
        \label{eq:tensor_decomp2}
\end{align}
$P^\mu$ is the average momentum, $P \equiv (p'+p)/2$, and $q$ is the momentum difference, $q \equiv p'-p$. The mass of meson $M$ is indicated by $m_M$, and its energy at momentum $\vec{p}$ is $E(p){=}\sqrt{m_M^2 + \vec{p}\,^2}$. To avoid complicated notation, we omit the index $M$ from the energy. Here we will use the notation $F^{M,f}_S \equiv A^{M^f}_{S10}\,,\,\,  F^{M,f}_V \equiv A^{M^f}_{10}\,,\,\,
F^{M,f}_T \equiv B^{M^f}_{T10}\,$. The decomposition simplifies in the rest frame, for which $\mathbf{p'} =\mathbf{0}$, that is
\begin{align}
    \langle M({p}')| \Op_S | M(p) \rangle &= \frac{1}{2 \sqrt{m\,E}}\,F^{M^q}_S \,, \\[1ex]
    \langle M(p') | \Op^0_V | M(p) \rangle &= \frac{(E+m)}{2 \sqrt{m\,E}}\, F^{M^q}_V\,, \\[1ex]
    \langle M({p}')| \Op^j_V | M(p) \rangle & = i\,\frac{ p^j}{2 \sqrt{m\,E(p)}}\, F^{M^q}_V\,, \\[1ex]
    \label{eq:tensor_decomp}
    \langle M({p}') | \Op^{0j}_T | M(p) \rangle &= i\, \frac{p^j}{2 m \sqrt{m\,E}} \,F^{M^q}_T\,,\\[1ex]
    \langle M({p}') | \Op^{jk}_T | M(p) \rangle & = 0\,,
\end{align}
where $E \equiv E(p)$, and $m\equiv m_M$. We note that the matrix element of Eq.~\eqref{eq:tensor_decomp2} is zero in the forward limit at any frame, due to the kinematic factor of $q$. Therefore, $F_T(Q^2=0)$ cannot be extracted directly from the lattice data.

\begin{figure}[h!]
    \centering
    \includegraphics[scale=0.24]{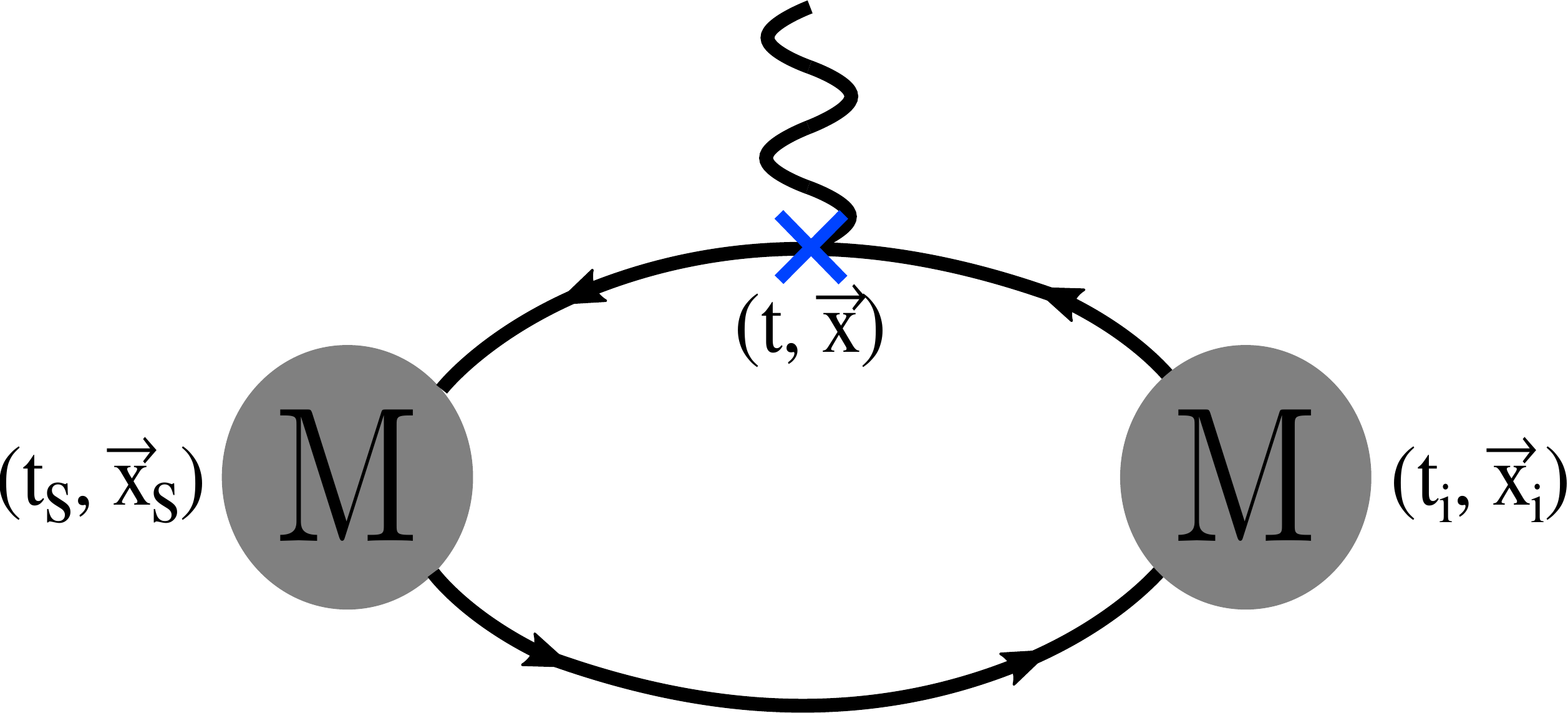}
\caption{Diagrammatic representation of  the three-point function entering the calculation of the connected contributions to the form factors. The wavy line indicates the current probing the meson.}
\label{fig:3ptDiagram}
\end{figure}

We calculate the connected contributions to the pion and kaon form factors, as shown schematically in Fig.~\ref{fig:3ptDiagram}. We use one ensemble of twisted-mass clover fermions and Iwasaki improved gluons labeled \texttt{cA211.30.32}. Besides the light mass-degenerate quarks, the ensemble has strange and charm quarks in the sea ($N_f=2+1+1$). These gauge configurations have been produced by the Extended Twisted Mass Collaboration (ETMC) and more details can be found in Ref.~\cite{Alexandrou:2018egz}. The most relevant ensemble parameters for this work are summarized in Table~\ref{tab:params}. Other characteristic parameters are $\kappa_\textrm{crit}=1/(2am_\textrm{crit}+8)=0.1400645$, $c_\mathrm{sw}=1.74$, $a\mu_{\ell}=0.003$,
$a\mu_\sigma=0.1408$, and $a\mu_\delta=0.1521$.
\begin{table}[h!]
\centering
\renewcommand{\arraystretch}{1.2}
\renewcommand{\tabcolsep}{6pt}
\begin{tabular}{| l| c | c | c | c | c  | c | c |}
    \hline
    \multicolumn{8}{|c|}{Parameters} \\
    \hline
    Ensemble   & $\beta$ & $a$ [fm] & volume $L^3\times T$ & $N_f$ & $m_\pi$ [MeV] &
    $L m_\pi$ & $L$ [fm]\\
    \hline
    cA211.30.32 & 1.726 & 0.09471(39)  & $32^3\times 64$  & 2+1+1 & 265
    & 4 & 3.0 \\
    \hline
\end{tabular}
\caption{Parameters of the ensemble used in this work. The lattice spacing is extracted from the pion sector~\cite{Alexandrou:2021gqw}.}
\label{tab:params}
\end{table}

We have a general setup for the calculation, and extract matrix elements in the rest, as well as the boosted frame. While the boosted frame is not necessary, it serves a very important purpose: it gives access to the form factors for a denser set of $Q^2$ because $Q^2$ is defined as follows in each frame:
\begin{eqnarray}
\label{eq:Q2_r}
Q_{\rm rest}^2 &=& 2 m (E(q) - m)\,,\\[1.5ex]
\label{eq:Q2_b}
Q_{\rm boosted}^2 &=& \mathbf{q}^2 - (E(p')-E(p))^2    \,.
\label{eq:Qboosted}
\end{eqnarray}
Another advantage of the boosted frame is that we obtain the form factors for an extended range of $Q^2$, up to 2.5 - 3 GeV$^2$. It is worth mentioning that, the signal-to-noise ratio for the rest frame at $Q^2=0$ remains constant~\cite{Lepage:1989hd} in the pion case. However, in the remaining of the cases, the matrix elements are subject to statistical fluctuations that cause a decrease in the quality of the signal.
Therefore, the use of the boosted frame, coupled with the momentum transfer, requires a larger number of statistics compared to the rest frame to control gauge noise. We use momentum boost of the form $\mathbf{p'}=2\pi \mathbf{n'}/L$ with $\mathbf{n'}=(\pm1,\pm1,\pm1)$, which results in an increase of the computational cost by a factor of eight. For $Q^2=0$ the eight combinations can be averaged, as we have done in the calculation for the Mellin moments  $\avgx$, $\avgxx$, and $\avgxxx$~\cite{Alexandrou:2020gxs,Alexandrou:2021mmi}. However, this is not the case for the form factors because the various $\mathbf{p'}$ do not correspond to the same value of $Q^2$, as can be seen in Eq.~(\ref{eq:Qboosted}). The choice of the momentum boost is such that it can be used to extract matrix elements with up to three-covariant derivative operators~\cite{Alexandrou:2021mmi} avoiding any mixing under renormalization.
The statistics used for each frame and each value of the sourse-sink time separation, $t_s$, is given in Tab.~\ref{tab:statistics}.

\begin{table}[h!]
\centering
\renewcommand{\arraystretch}{1.2}
\renewcommand{\tabcolsep}{6pt}
\begin{tabular}{| c | c | c | c | c | c |}
    \hline
    $\mathbf{n'}$  &  $t_s/a$  & $N_{\rm confs}$   & $N_{\rm src}$  & $N_{p'}$  & Total statistics \\
    \hline
  (0,0,0)  &  12,  14, 16, 18, 20, 24 &  122 & 16 & 1 & 1,952 \\
  $(\pm1,\pm1,\pm1)$  &  12 & 122 & 48 & 8 & 46,848 \\
 $(\pm1,\pm1,\pm1)$   &   14, 16, 18  &   122 & 104 & 8 & 101,504 \\
    \hline
\end{tabular}
\caption{Statistics used in the calculation of the form factors in the rest and boosted frames. $t_s$ is the source-sink time separation. $N_{\rm confs}$ is the number of configurations, $N_{\rm src}$ is the number of source positions per configuration, and $N_{p'}$ is the number of combinations for $\mathbf{p'}$.}
\label{tab:statistics}
\end{table}

\section{Renormalization}
\label{sec:renorm}

We apply renormalization functions that are calculated non-perturbatively using the Rome-Southampton method (RI$'$ scheme)~\cite{Martinelli:1994ty} following the procedure outlined in Refs.~\cite{Alexandrou:2010me,Alexandrou:2012mt,Alexandrou:2015sea}. We employ the momentum source approach, introduced in Ref.~\cite{Gockeler:1998ye}, in which the vertex functions are calculated with a momentum-dependent source. The momentum-source method requires separate inversions for each value of the renormalization scale, but has the advantage of high statistical accuracy and the evaluation of the vertex for any operator at negligible computational cost. The renormalization functions in the RI$'$ scheme are defined by the condition
\bea
\label{renormalization cond}
   {\cal Z}_q^{-1}\,{\cal Z}_{\cal O}\,\frac{1}{12} {\rm Tr} \left[\Gamma_{\cal O}^L(p)
     \,\left(\Gamma_{\cal O}^{{\rm tree}}(p)\right)^{-1}\right] \Bigr|_{p^2=\mu_0^2} = 1\, ,
\eea
where $\Gamma^L_{\cal O}$ is the amputated vertex function of operator ${\cal O}$, and $\Gamma^{\rm tree}_{\cal O}$ is its tree-level value. Here, we focus on the renormalization for the scalar and tensor form factors. For the vector form factor we use the conserved current, which is free of renormalization. Note that, for twisted mass fermions, the scalar form factor is renormalized using the vertex functions of the pseudoscalar operator. The latter suffers from a pion pole and a dedicated analysis is needed to extract it reliably. For the ensemble used in this work, we use the $Z_p$ results given in Ref.~\cite{Alexandrou:2021gqw}. The momentum of the vertex function in Eq.~\eqref{renormalization cond} is indicated by $p$, and is set to the RI$'$ renormalization scale, $\mu_0$. ${\cal Z}_q$ is the renormalization function of the fermion field defined as
\bea
   {\cal Z}_q = \frac{1}{12} {\rm Tr} \left[(S^L(p))^{-1}\, S^{{\rm tree}}(p)\right] \Bigr|_{p^2=\mu_0^2}\,,
\eea
where $S^L(p)$ ($S^{{\rm tree}}$) is the lattice (tree-level) fermion propagator. In the above equations,  ${\cal Z}_q \equiv {\cal Z}_q^{\rm RI'}(\mu_0, m_\pi)$ and
${\cal Z}_{\cal O} \equiv {\cal Z}_{\cal O}^{\rm RI'}(\mu_0, m_\pi)$, where $m_\pi$ is the pion mass of the ensemble used. 

We apply a chiral extrapolation on ${\cal Z}_T$ calculated on a set of ensembles with different pion mass (see Table~\ref{tab:Z_ensembles}), and we find a negligible pion mass dependence. Therefore, the linear fit with respect to $m_\pi^2$ (or, equivalently linear in the twisted mass parameter),
\be
{\cal Z}_{\cal O}^{\rm RI'}(\mu_0, m_\pi) = {\cal Z}^{{\rm RI'}, \rm chiral}_{\cal O} (\mu_0) + {\cal Z}_{\cal O}^{(1)} (\mu_0)\cdot(a\,m_\pi)^2\,,
\ee
yields a zero slope. We convert the chirally extrapolated $Z_P$ and $Z_T$ to the $\overline{\rm MS}$ scheme and evolve them to a scale of 2 GeV using an intermediate Renormalization Group Invariant (RGI) scheme. The factor $C^{\overline{\rm MS},{\rm RI'}}$ needed for the conversion
\be
{\cal Z}_{\cal O}^{\overline{\rm MS}}(2\, {\rm GeV},\mu_0) =  C_{\cal O}^{\overline{\rm MS},{\rm RI'}}(2\, {\rm GeV},\mu_0) \,\cdot\, {\cal Z}_{\cal O}^{{\rm RI'}, \rm chiral}(\mu_0)\,,
\ee
is calculated up to four loops (three loops) in perturbation theory for the pseudoscalar (tensor) operator~\cite{Chetyrkin:1997dh,Vermaseren:1997fq,Gracey:2000am,Gracey:2003yr}.
Finally, we apply a linear fit in $(a \mu_0)^2$  to eliminate residual dependence on the initial scale $\mu_0$, that is
\begin{equation}
{\cal Z}_{\cal O}^{\overline{\rm MS}}(2\, {\rm GeV},\mu_0)= Z^{\overline{\rm MS}}_{\cal O}(2\, {\rm GeV}) + Z_{\cal O}^{(1)}\cdot(a\,\mu_0)^2\,.
\label{Zfinal}
\end{equation}
$Z^{\overline{\rm MS}}_{\cal O}$ corresponds to the final value of the renormalization function for operator ${\cal O}$. 

\begin{table}[!h]
\begin{center}
\begin{tabular}{ccc}
\hline
\hline
$\,\,\,$        $\,\,\,$              & $\beta=1.726$, $\,\,\,a=0.093$ fm  &          \\ \hline
\hline
$a \mu$   & $a m_{PS}$ & lattice size\\
\hline
$\,\,\,$  0.0060$\,\,\,$         & 0.1680   & $24^3 \times 48$ \\  
$\,\,\,$  0.0080$\,\,\,$         & 0.1916   & $24^3 \times 48$ \\          
$\,\,\,$  0.0100$\,\,\,$         & 0.2129   & $24^3 \times 48$ \\  
$\,\,\,$  0.0115 $\,\,\,$        & 0.2293   & $24^3 \times 48$ \\ 
$\,\,\,$  0.0130 $\,\,\,$        & 0.2432   & $24^3 \times 48$  \\ 
\hline
\hline
\end{tabular}
\caption{Parameters for the $N_f=4$ ensembles used for the renormalization functions.} 
\label{tab:Z_ensembles}
\end{center}
\end{table}

We evaluate the renormalization functions on the five ensembles of Table~\ref{tab:Z_ensembles} for a wide range of values for $(a\,p)^2$ that are spatially isotropic, that is
\begin{equation}
\label{eq:P4}
(a\, p) \equiv 2\pi \left(\frac{n_t}{L_t}+\frac{1}{2\,L_t},
\frac{n_x}{L_s},\frac{n_x}{L_s},\frac{n_x}{L_s}\right)\,,  \qquad\quad n_t \,\epsilon\, [2, 9]\,,\quad n_x\,\epsilon\, [2, 5]\,,\quad (a p)^2 \in [0.9 - 6.7]\,,
\end{equation} 
where $L_t$ ($L_s$) is the temporal (spatial) extent of the lattice. The choice of momenta is such that the ratio $ {\sum_i p_i^4}/{(\sum_i p_i^2 )^2}$ is less than 0.3, to reduce Lorentz non-invariant contributions~\cite{Constantinou:2010gr}. Such terms have been found in the analytic expressions from lattice perturbation theory~\cite{Constantinou:2009tr,Alexandrou:2010me,Alexandrou:2012mt,Constantinou:2013ada}. We improve ${\cal Z}_{\cal O}$ by subtracting inherited lattice artifacts utilizing perturbation theory. More precisely, we calculate in one-loop perturbation theory the terms to all orders in the lattice spacing, ${\cal O}(g^2\,a^\infty)$~\cite{Constantinou:2014fka,Alexandrou:2015sea}. These are subtracted from the final estimates of ${\cal Z}_{\cal O}$. 

\begin{figure}[!h]
\centerline{\includegraphics[scale=0.39]{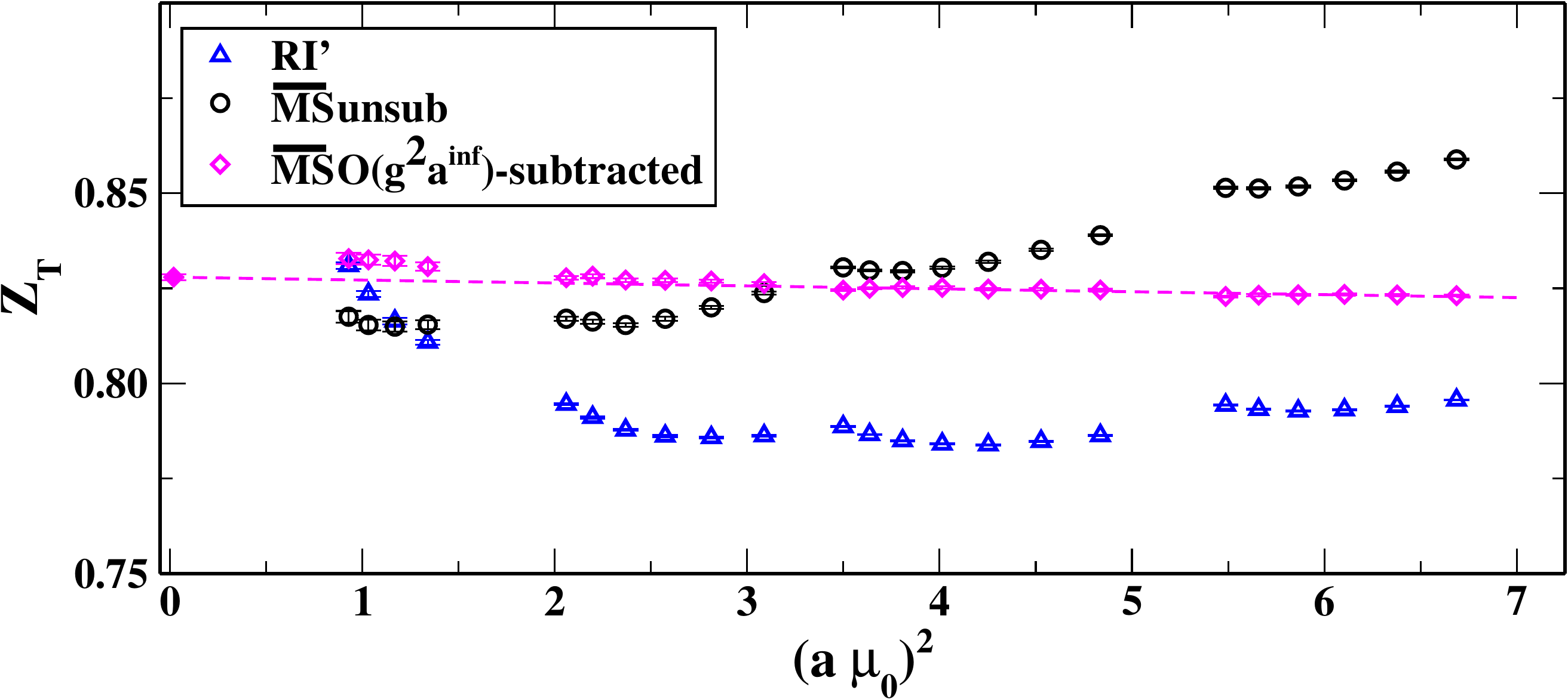}}
\vspace*{-0.2cm}
\caption{Chirally extrapolated estimates for $Z_{\rm T}$. Results in the RI$'$ scheme are shown with blue triangles, and in the ${\overline{\rm MS}}$ scheme at 2 GeV with black circles. The magenta diamonds correspond to the improved estimates upon subtraction of the artifacts applied on the black points. The dashed line corresponds to the fit of Eq.~(\ref{Zfinal}). The filled magenta diamond is our final value.}
\label{ZRIMS}
\end{figure}

The renormalization function $Z_T$ is shown in Fig.~\ref{ZRIMS} as a function of the initial scale $\mu_0$. $Z_P$ can be found in Ref.~\cite{Alexandrou:2021gqw}. For $Z_T$, the conversion to the ${\overline{\rm MS}}$ and evolution to a common scale results in a flatter behavior in the initial scale, as compared to the RI$'$ estimates. This effect is more profound in the region $(a\mu_0)^2 < 2$, where hadronic contamination are non-negligible. However, as $(a\mu_0)^2$ increases, a slope is observed in the pure non-perturbative estimates due to lattice artifacts. The subtraction procedure using the one-loop perturbative expressions removes the majority of the artifacts leading to plateaus with negligible slope. 

\noindent
For $Z_{\rm T}$ we apply the fit of Eq.~(\ref{Zfinal}) in the interval $(a\,\mu_0)^2\, \epsilon\, [2-7]$, where plateau is identified. We obtain 
\begin{eqnarray}
Z^{\overline{\rm MS}}_{\rm T} (2\, {\rm GeV}) &=& 0.829(1) \,, \\[0.5ex]
Z^{\overline{\rm MS}}_{\rm P} (2\, {\rm GeV}) &=& 0.475(4)\,.
\end{eqnarray}
For $Z_P$ we use the average of the results presented in Ref.~\cite{Alexandrou:2021gqw}.

\section{Extraction of matrix elements}
\label{sec:analysis}

For the extraction of the matrix elements $\langle M({p}') | \Op_\Gamma | M({p})$ we use the interpolating fields of $\pi^+$ and $K^+$
\begin{eqnarray}
J_{\pi^+} = \overline{d}\gamma_5 u\,, \qquad
J_{K^+} = \overline{s}\gamma_5 u\,,
\end{eqnarray}
which are smeared at the source and the sink using Gaussian smearing. Further details can be found in Ref.~\cite{Alexandrou:2020gxs}. 
The matrix elements are obtained from a combination of two-point functions,
\begin{equation}
    C_M(t,\mathbf{p})=\sum_{\mathbf{x}} \langle 0 | J_M(t,\mathbf{x})J_M^\dag (0,\mathbf{0})  | 0 \rangle e^{i\mathbf{p}\cdot\mathbf{x}}\,,
\end{equation}
and three-point functions 
\begin{equation}
\label{eq:C3pt}
C_M^{\Gamma}(t,t_s,\vec{p}) = \sum_{\vec{x}_s,\vec{x}} \langle 0 | J_M(t_s,\vec{x}_s){\cal O}_\Gamma(t,\vec{x})J^\dag_M(t_i,\vec{x}_i) |  0 \rangle  e^{-i\vec{p}\cdot(\vec{x}_s-\vec{x}_i)}\,.
\end{equation} 
In the above equations, $t_i$, $t$, and $t_s$ denote the source, insertion, and sink Euclidean times, respectively. The spatial coordinates of the source, current insertion and sink are $\vec{x}_i$, $\vec{x}$, and $\vec{x}_s$. Without loss of generality, we set the source to be at $t_i=0$, so that the source-sink separation is $t_s -t_i \equiv t_s$. 
The normalization of the meson state is $\langle 0 |J_M|M\rangle = Z_M/\sqrt{2 E}$. In the results presented in this paper, we focus on the $u^+$ contribution to the pion, where $q^+\equiv q+\bar{q}$. 

\begin{figure}[h]
    \centering
    \includegraphics[scale=0.26]{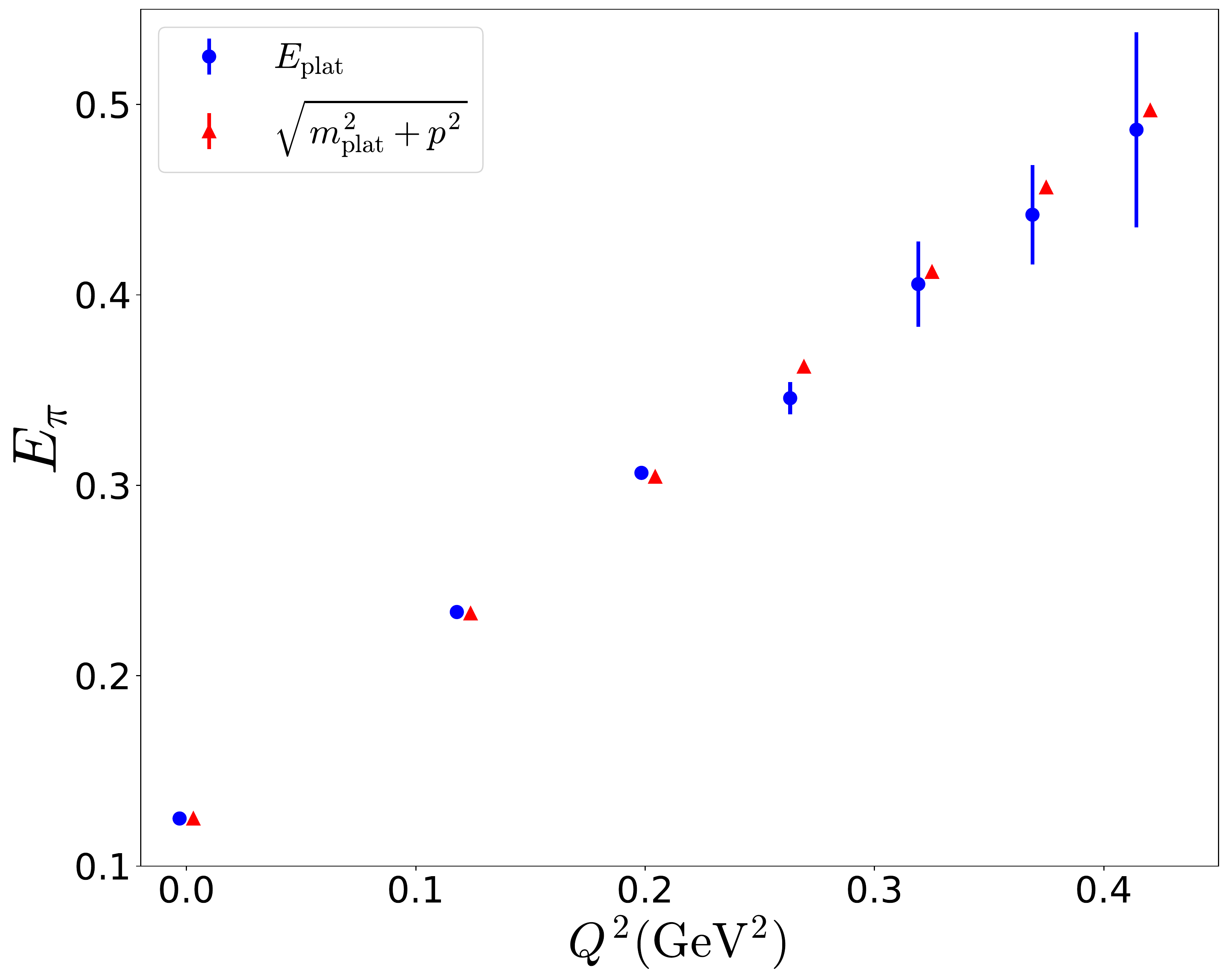}
    \includegraphics[scale=0.26]{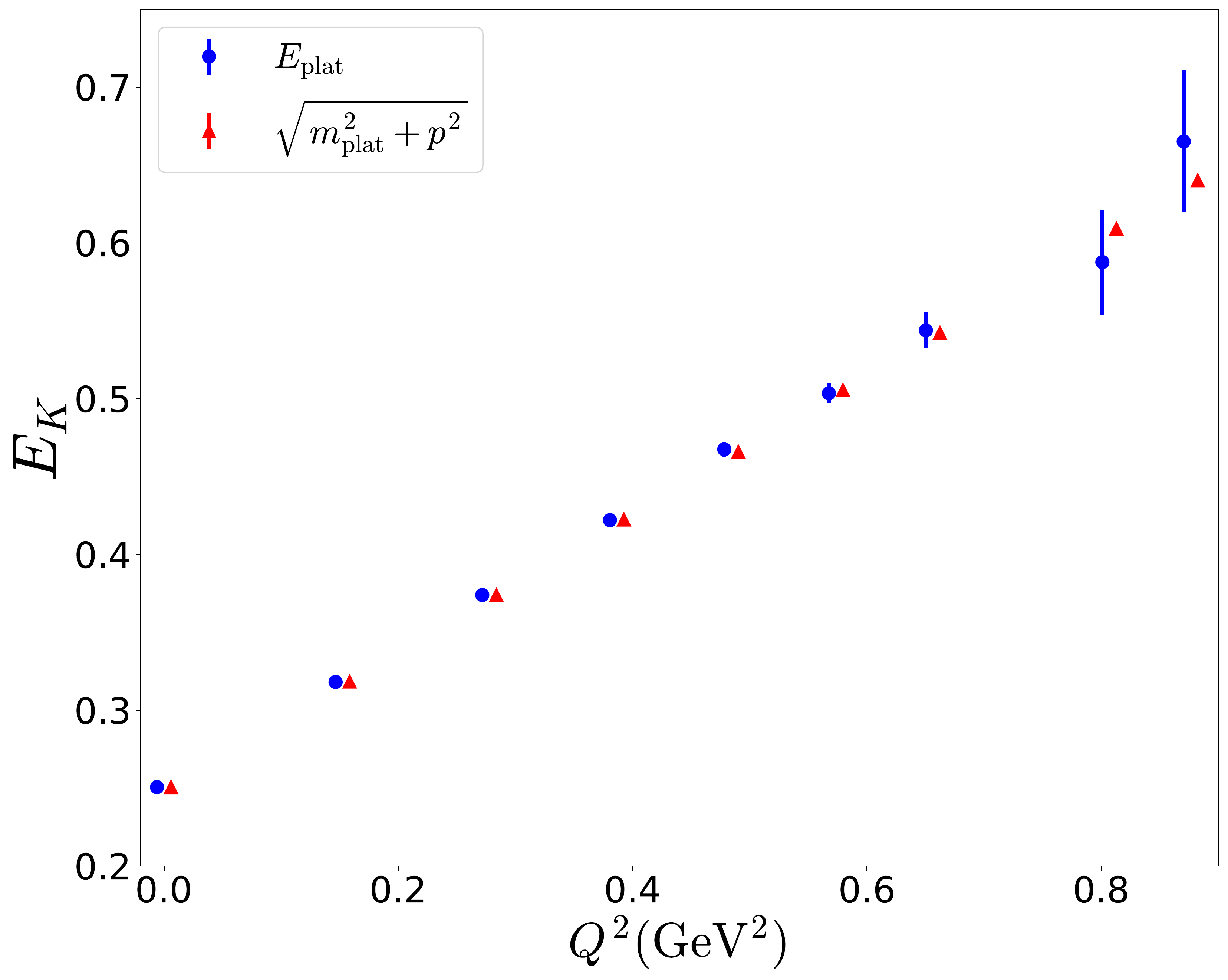}
    \caption{The ground state energies for the pion (left) and kaon (right) at the various values of $Q^2$. The red points correspond to the dispersion relation, and the blue to the energy obtained from a plateau fit. The points are slightly shifted horizontally for clarity.}
    \label{fig:dispersion_relation_test_pion_rest}
\end{figure}

To extract the meson matrix elements, we form the optimized ratio 
\begin{equation}
    \label{eq:ratio}
    R^M_\Gamma(t_s,t;\mathbf{p}', \mathbf{q} = \mathbf{p}' - \mathbf{p}) = \frac{ C^M_\Gamma(t_s,t;\mathbf{p}',\mathbf{q}) }{ C_M(t;\mathbf{p}'^2) }
    \sqrt{ \frac{ C_M(t_s-t;\mathbf{p}^2) C_M(t;\mathbf{p}'^2) C_M(;t_s\mathbf{p}'^2) }{ C_M(t_s-t;\mathbf{p}'^2) C_M(t;\mathbf{p}^2) C_M(t_s;\mathbf{p}^2) } } \,,
\end{equation}
which cancels the time dependence  and the overlaps between the interpolating field and the meson state. We note that the ratio of Eq.~\eqref{eq:ratio} is written for a general frame. We use $C_M(t;\mathbf{p}^2)=c_0 e^{-E_0(\mathbf{p}^2) t}$ for the two-point functions, where $c_0$ is calculated from the two-state fit on the two-point functions\alx{,} and $E_0=\sqrt{m^2 + \mathbf{p}^2}$ is calculated from the plateau fit on the effective mass, so that, at insertion times far enough away from the source and sink positions, the ratio becomes independent of the insertion time, i.e., 
\begin{equation}
    R^M_\Gamma(t_s,t;\mathbf{p}', \mathbf{q}) \xlongrightarrow[\text{$Et\gg1$}]{\text{$\Delta E(t_s-t)\gg1$}}  \Pi^M_\Gamma(t_s; \mathbf{p}', \mathbf{q}) \,.
    \label{eq:ratio2}
\end{equation}
We calculate $\Pi^M_\Gamma$ using two methods: \textbf{(a)} by fitting the plateau region of the data to a constant value; \textbf{(b)} by performing a two-state fit on the three-point functions. We use the general Ansatz
\begin{align}
\begin{split}
    C^M_\Gamma(t_s, t; \mathbf{p}', \mathbf{p} = \mathbf{p}' - \mathbf{q}) = &A_{00} e^{ -E_0'(\mathbf{p}'^2) (t_s - t_i) - E_0(\mathbf{p}^2) t_i }
    + A_{01}e^{ -E_0'(\mathbf{p}'^2) ( t_s - t_i ) - E_1(\mathbf{p}^2) t_i } \\
    &+ A_{10} e^{ -E_1'(\mathbf{p}'^2) ( t_s - t_i ) - E_0(\mathbf{p}^2) t_i }
    + A_{11} e^{ -E_1'(\mathbf{p}'^2) ( t_s - ti ) - E_1(\mathbf{p}^2) t_i } \,,
    \label{eq:two_state_fit_Boosted}
\end{split}
\end{align}
which simplifies in the rest frame ($\vec{p}'=0$, $E_0=m_0$, $E_1=m_1$).
In the above expressions, $m_0$ ($m_1$) are the masses of the ground (first-excited) state, and $E_0$ ($E_1$) its corresponding energy. The masses are fixed from the fit on the two-point functions. For the two-state fit, the time-independent ratio is then $\Pi^M_\Gamma=A_{00} / \sqrt{c_0' c_0}$, where $c_0$ ($c_0'$) is the amplitude calculated from the two-state fit on the two-point functions at the initial (final) momentum. To improve the quality of the signal for the two-state fit, we use the dispersion relation for the ground-state energy, $E_0(\mathbf{p}^2) = \sqrt{m^2 + \mathbf{p}^2}$, while $E_1$ is calculated from the two-state fit on the two-point functions. In the boosted frame, we obtain the ratio of Eq.~\eqref{eq:ratio} for the cases where the momentum of the two-point function is less or equal to $12\frac{2\pi}{L}$. We find that for higher values of the momentum, the data are very noisy and subjected to systematic uncertainties. To test the validity of this approach, we compare the ground-state energy calculated by the dispersion relation to those extracted from plateau fits of the effective energies at each value of the momentum transfer. Figure~\ref{fig:dispersion_relation_test_pion_rest} shows this comparison for the pion and kaon. We find that the dispersion relation holds, and suppresses statistical uncertainties.

For each value of the 4-vector momentum squared, $Q^2=(E' - E)^2 + (\mathbf{p}' - \mathbf{p})^2$, there are more than one combination of $\mathbf{p}'$ and $\mathbf{p}$ contributing. Also, the vector and tensor form factors may be extracted from a current insertion with different Dirac matrices ($\mu,\,\nu=0,1,2,3$). Therefore, it is important to calculate the statistical uncertainties appropriately. This is particularly needed in the boosted frame, as one combines two-point functions at different momenta in the ratio of Eq.~(\ref{eq:ratio}). For each value of $Q^2$ we solve the system of equations 
\begin{equation}
    \Pi_\Gamma(\mathbf{p}',\mathbf{q}) = G(\mathbf{p}',\mathbf{q}) F(Q^2) \,,
    \label{eq:decomp}
\end{equation}
where $G$ is a vector of kinematic coefficients given by the decomposition, and $F$ is a vector of the form factors. Because $\Pi$ and $G$ depend on the momentum vectors $\mathbf{p}'$ and $\mathbf{q}$ but $F$ depends on the four-momentum $Q^2$, the system is over constrained and so we use singular value decomposition (SVD). Since for the matrix elements that we study here there is only one form factor\del{s}, the above procedure is equivalent to a weighted average $\Pi/G$ over the values of $\mathbf{p}'$ and $\mathbf{q}$ which contribute to the same $Q^2$. 

\section{Pion Form Factors}
\label{sec_pion}

\subsection{Excited-states investigation}

As previously mentioned, we calculate the form factors in both the rest ($\mathbf{p'=0}$) and boosted ($\mathbf{p'\ne 0}$) frame. The analysis in the rest frame is rather straightforward, because the plus/minus components of the momentum transfer vector $\mathbf{q=p'-p}$ contribute to the same $Q^2$, and one can take their average, weighted by their jackknife errors. 
Furthermore, the square root in the ratio of Eq.~(\ref{eq:ratio}) only receives contributions from the two-point function at zero momentum and at the momentum of the source. Because of the aforementioned properties, the form factors have small statistical uncertainties. 

\begin{figure}[h!]
    \centering
    \includegraphics[scale=0.3]{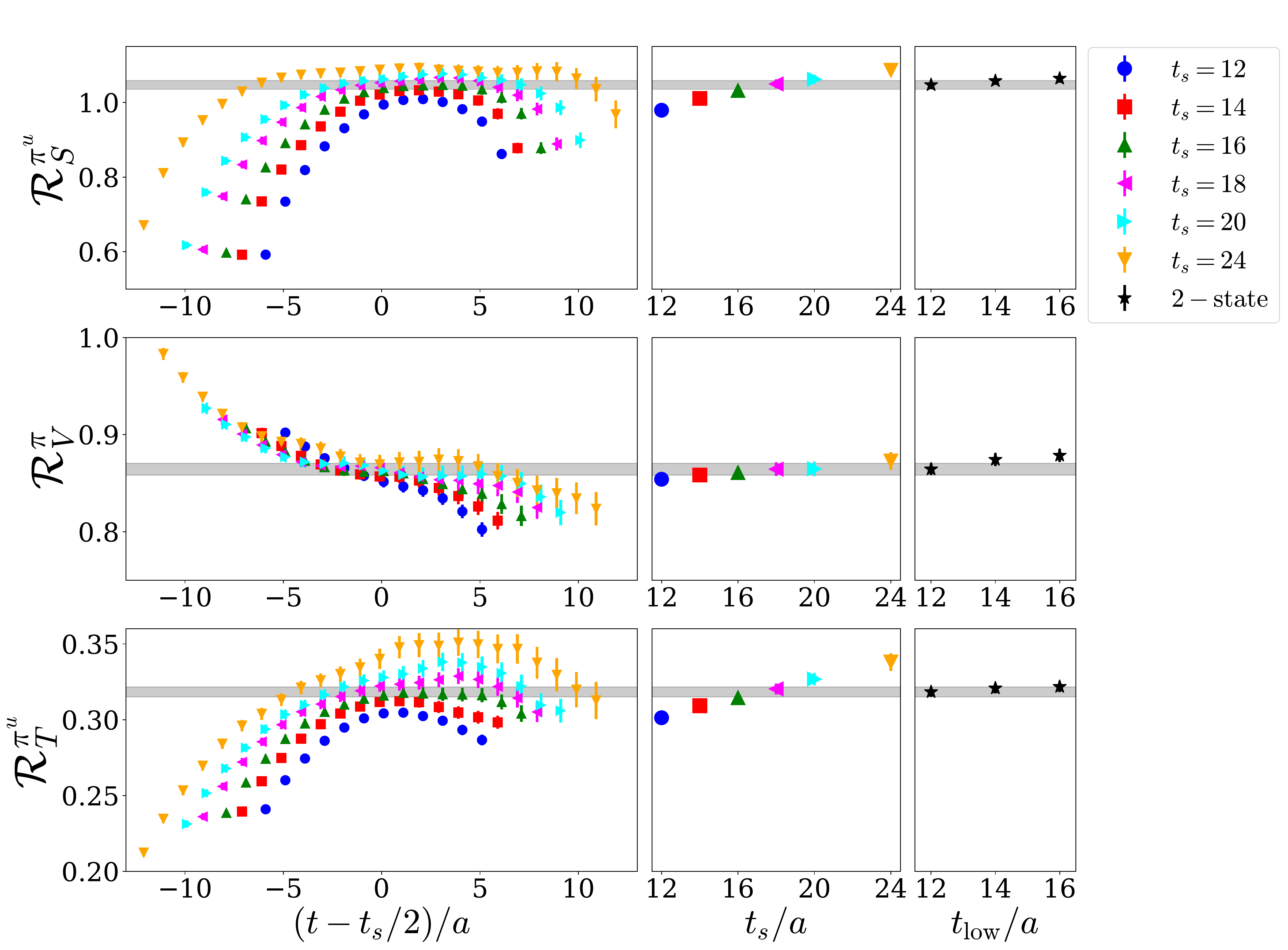}
    \caption{From top to bottom: ratio of the scalar, vector and tensor pion form factor as a function of the rescaled current insertion time, $t/a-t_s/2a$, for the first nonzero momentum transfer, $Q^2\sim0.12\,{\rm GeV}^2$. The plateau fits are plotted in the center panels as a function of $t_s$ and. The right panels show the two-state fit results by varying the lowest $t_s$ entering the fit. In all panels, the two-state fit is plotted in all panels as a gray band. }
    \label{fig:Ratio_pion_rest}
\end{figure}

In Fig.~\ref{fig:Ratio_pion_rest} we plot the ratio of Eq.~(\ref{eq:ratio}) for the three operators, as obtained in the rest frame at $Q^2\sim0.12\,{\rm GeV}^2$. We find that excited states affect the scalar and tensor operators. The vector one, at this momentum transfer, seems to have small contamination from excited states. In all cases, the two-state results are compatible with the plateau fits for $t_s\ge 18a$, with borderline agreement at $t_s=16 a$. We also find very little sensitivity in the lowest value of $t_s$ entering the fit, as can be seen in the right panel of Fig.~\ref{fig:Ratio_pion_rest}.

\begin{figure}[h!]
    \centering
    \includegraphics[scale=0.28]{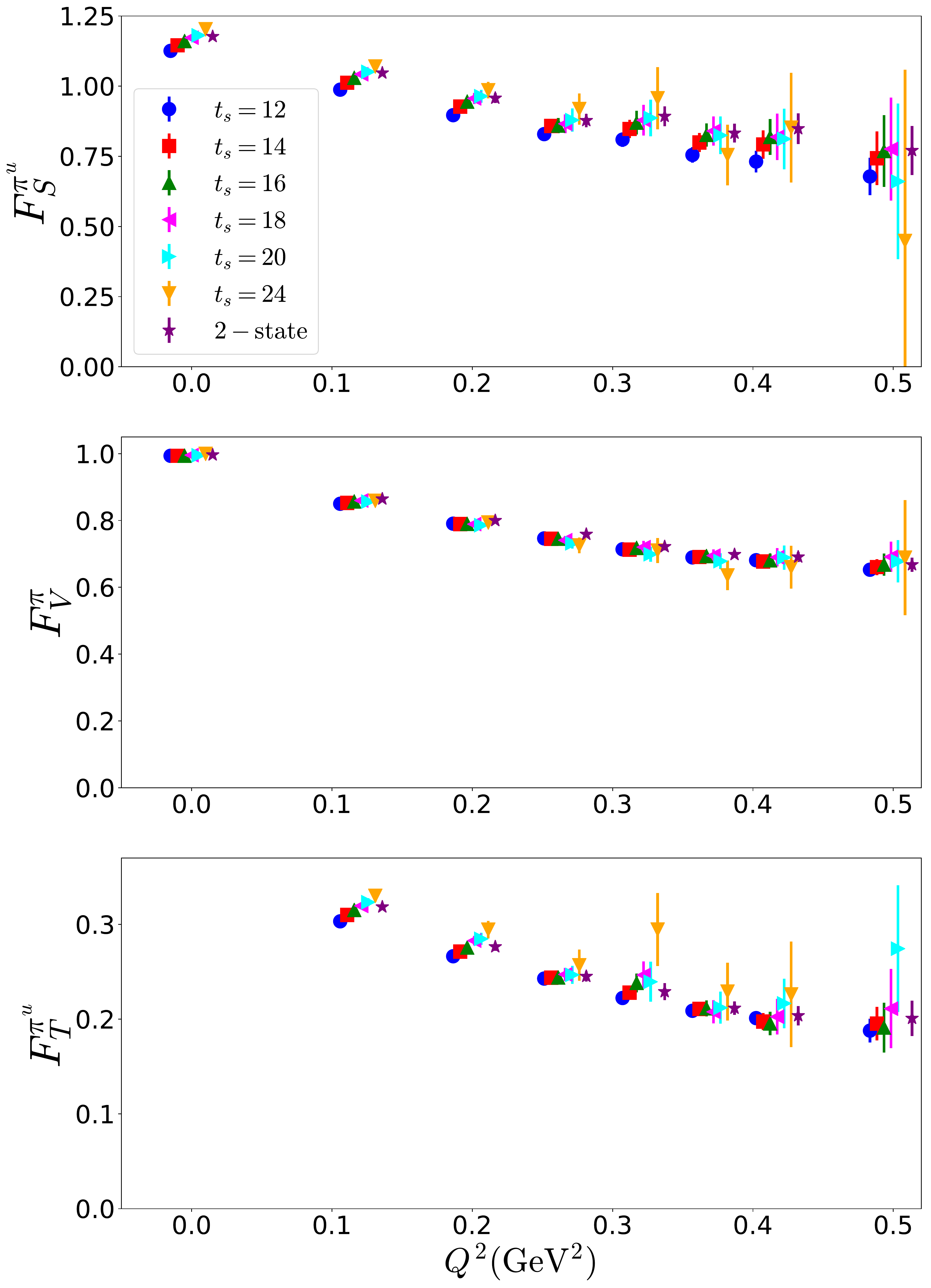}
    \caption{From top to bottom: The scalar, vector and tensor form factors in the rest frame for $t_s/a=$ 12 (blue circles), 14 (red squares), 16 (green up triangles), 18 (magenta left triangles), 20 (cyan right triangles), 24 (orange down triangles). The two-state fit on these data is shown with purple stars.}
    \label{fig:FFs_pion_rest}
\end{figure}

\vspace*{1cm}
The $Q^2$ dependence of the scalar, vector and tensor form factors using the data in the rest frame is shown in Fig.~\ref{fig:FFs_pion_rest}. We include the data for all values of $t_s$, that is, $12a$ - $24a$, as well as the 2-state fits using the aforementioned separations. We find good signal for $Q^2$ up to $0.5$ GeV$^2$. This is expected as the pion is a light particle, and the ratio $|Q_{\rm max}|/m_\pi$ is about 2.5 for this ensemble. 

The tensor form factor is suppressed by the quark mass due to the chirality flip on a quark line, which explains the fact that our results are about 30$\%$ of the vector one. Also, $F^{\pi^u}_T(Q^2=0)$ cannot be extracted directly from the matrix elements, due to a vanishing kinematic factor (see Eq.~\eqref{eq:tensor_decomp}). Therefore, one has to fit the $Q^2$ dependence of the tensor form factor to extract its forward limit. $F^{\pi^u}_T(Q^2=0)$ is an important quantity, as it defines the tensor anomalous magnetic moment $\kappa_T$~\cite{Burkardt:2005hp}.

Regarding excited-states effect, the vector form factor converges to the ground state at source-sink time separations about 1.3 fm ($t_s=14a$) for all values of $Q^2$. On the contrary, the scalar and tensor form factors suffer from excited-states effect, and the ground state contribution is identified at about 1.7 fm ($t_s=18a$) and higher. This conclusion is based on the agreement between the single-state fits results for the form factors at these separation and the two-state fit values. For the scalar, the excited-states effect are visible for the whole range of $Q^2$, even though the statistical uncertainties increase. Such a behavior is expected for the scalar operator, which is known to suffer from large excited-states contamination also in the nucleon case (see, e.g., Ref.~\cite{Alexandrou:2019brg}). The excited-states contamination for the tensor form factor decrease with increase of $Q^2$.
Comparison of the form factors, for example at $t_s=18 a$ reveals that the statistical uncertainties for the vector are the smallest, and for the scalar the largest. At $Q^2\sim 0.5$ GeV$^2$, the relative statistical errors of the tensor and scalar form factor are 3 and 3.5 times higher than the vector. 

\begin{figure}[h!]
\hspace*{-0.75cm}
    \includegraphics[scale=0.25]{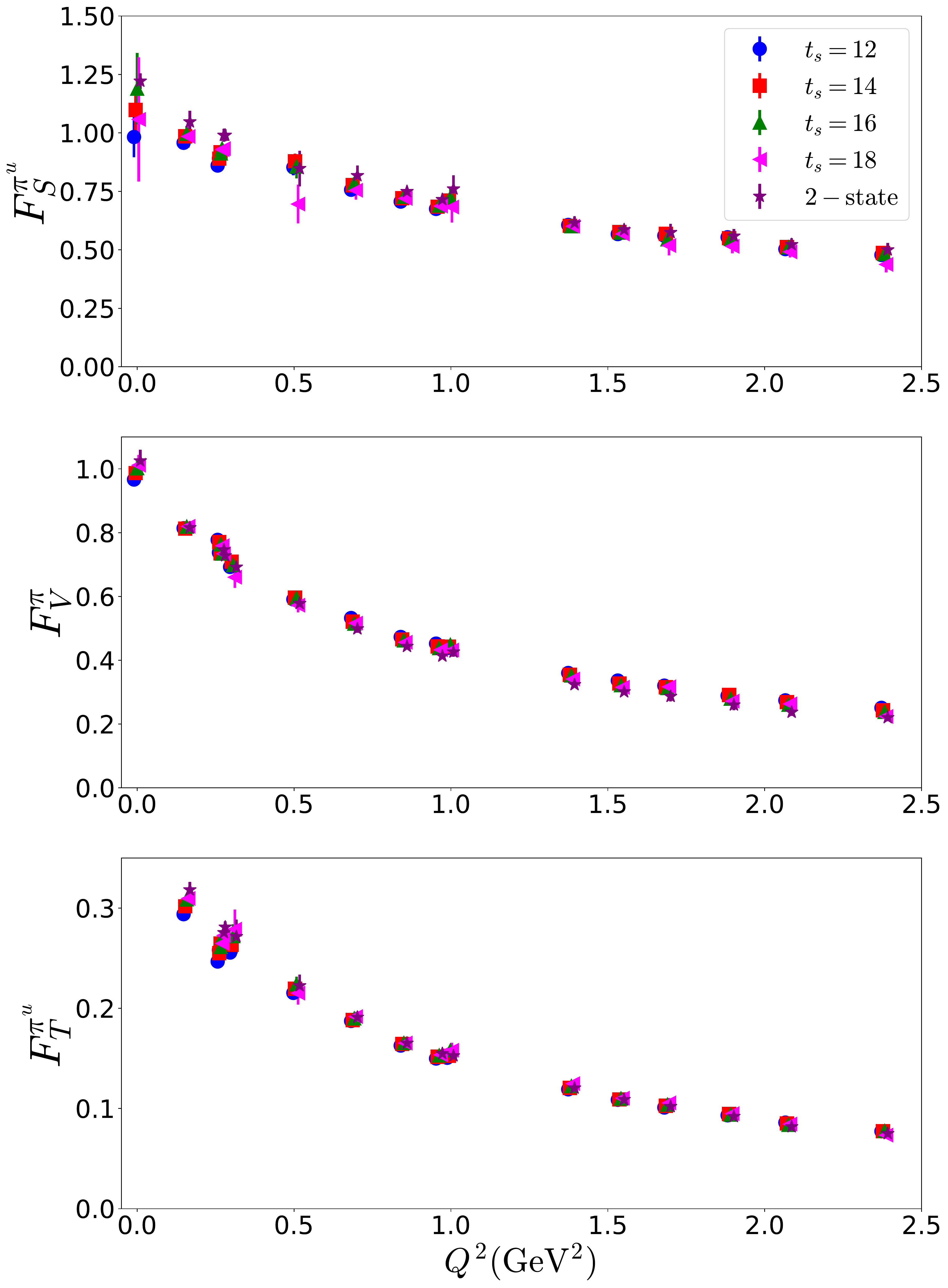}\hspace*{-0.35cm}
        \includegraphics[scale=0.25]{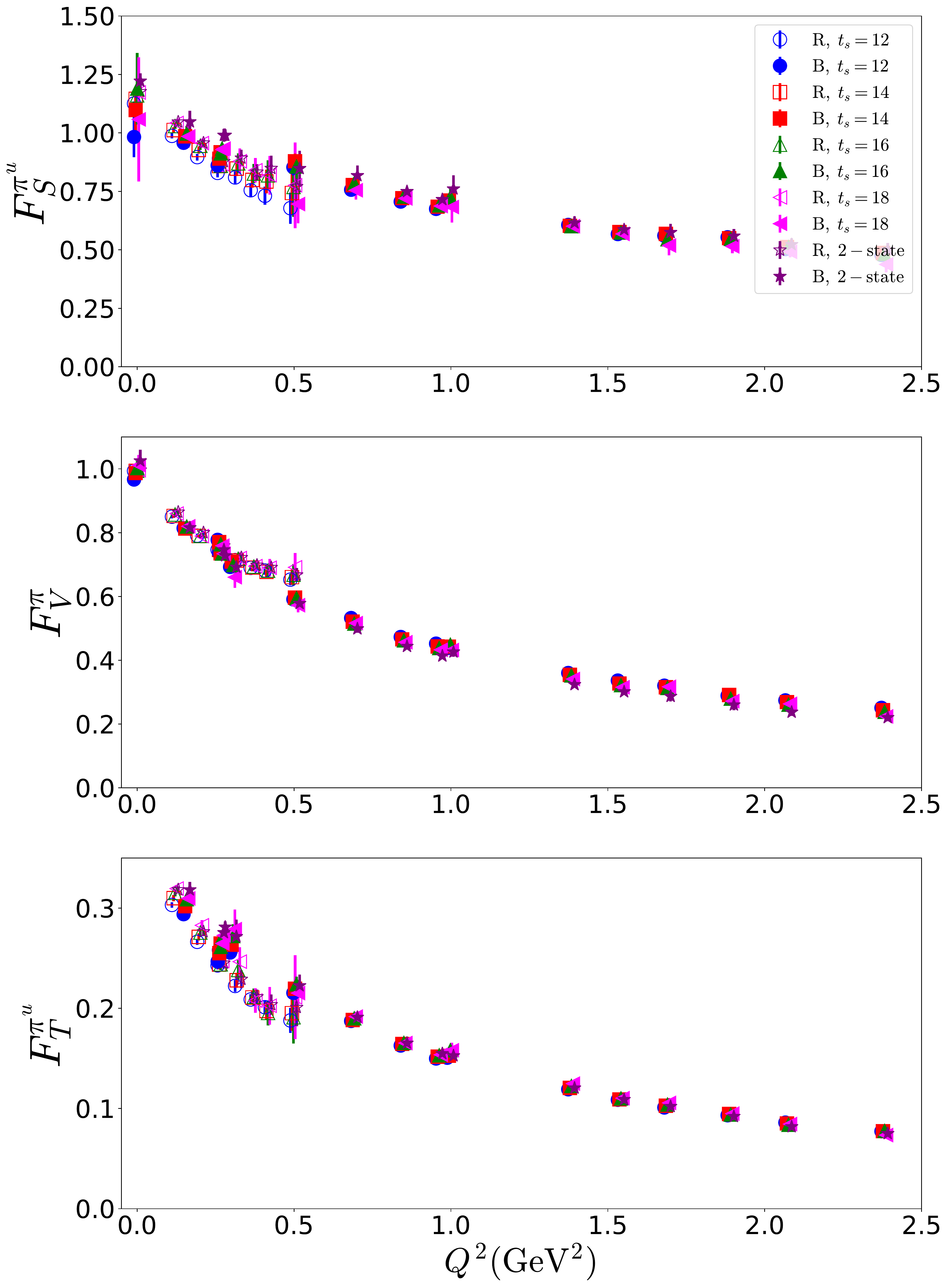}
    \caption{Left from top to bottom: The scalar, vector and tensor form factors in the boosted frame for $t_s/a=$ 12 (blue circles), 14 (red squares), 16 (green up triangles) and 18 (magenta left triangle). The two-state fit as applied on these data is shown with purple stars. Right from top to bottom: Comparison of the scalar, vector and tensor form factors in the rest (open symbols) and boosted (filled symbols) frame. Blue, red, green, and magenta points correspond to $t_s/a=12,\,14,\,16,\,18$. }
    \label{fig:FFs_pion_boost}
\end{figure}

The meson states for the boosted frame setup, carry momentum 0.72 GeV, leading to a ground state energy of about 0.77 GeV. We obtain a good signal up to $Q^2\sim 2.5$ GeV$^2$, which corresponds to a ratio $|Q_{\rm max}|/E_0 \sim 2$, similar to the ratio in the rest frame. As we have done in our previous work on the pion and kaon Mellin moments $\avgx$,  $\avgxx$, and $\avgxxx$~\cite{Alexandrou:2020gxs,Alexandrou:2021mmi}, we focus on four values of the source-sink time separation, that is $t_s/a=12,\,14,\,16,\,18$. We have shown that at $Q^2=0$ a two-state fit on these separations can eliminate excited states. Our results shown in the left panel of Fig.~\ref{fig:FFs_pion_boost} indicate that the excited-states for the vector are present only at $Q^2=0$. For the tensor and scalar form factors the excited states are suppressed starting at $Q^2 \sim 0.5$ GeV$^2$ and $Q^2 \sim 1$ GeV$^2$, respectively.

It is interesting to compare the form factors as extracted from the rest and boosted frame. Since there is no frame dependence in these quantities, the lattice data in the two frames should be compatible. However, systematic uncertainties, such as cutoff effects, may affect each frame differently. In principle, combining the form factors for the two frames gives a more dense set of data for $Q^2\le0.4$. This is useful, as one can control better the parametrizations of the $Q^2$ dependence (see, Sec.~\ref{sec:fits_pion}).  In the right panel of Fig.~\ref{fig:FFs_pion_boost} we show the comparison, and one can see that there is very good agreement between the two frames for the vector. For the scalar, the same conclusion holds at a separation where ground-state dominance has been established, and for the two-state fits data. Finally, the tensor exhibits some tension in the slope between $0.25$ GeV$^2$ and $0.5$ GeV$^2$. In the remaining of the analysis, we will use the 2-state fit results and combine both the rest and boosted frame.

\subsection{Parametrization of pion form factors}
\label{sec:fits_pion}

To extract the pole parameters from the form factors, we parametrize their $Q^2$ dependence using the monopole Ansatz as depicted by the Vector Meson Dominance (VMD)
model~\cite{OConnell:1995fwv},
\begin{equation}
\label{eq:fit}
F_\Gamma(Q^2) = \frac{F_\Gamma(0)}{1 + \frac{Q^2}{{\cal M}_\Gamma^2}} \,,  
\end{equation}
with two fit parameters, that is the forward limit of the form factor, $F_\Gamma(0)$, and the monopole mass, ${\cal M}_\Gamma$. This fit function describes better the $Q^2$-dependence of the lattice data for all operators than a dipole fit. Note that, while such an Ansatz is commonly used, it does not rely on theoretical arguments. For the scalar and vector form factors we also apply a one-parameter fit by fixing $F_\Gamma(0)$ to the value obtained from our lattice data. 

Several interesting quantities can be derived from parametrizing the form factors. The most commonly extracted quantity is the radius, defined as the slope of the form factor at $Q^2=0$
\begin{equation}
\langle r^2 \rangle_\Gamma = -\frac{6}{F_\Gamma(0)} \frac{\partial F_\Gamma (Q^2)}{\partial Q^2}\Bigg{|}_{Q^2=0}\,.  
\end{equation}
For the ansatz of Eq.~\eqref{eq:fit} for the form factors, the radius is related to the monopole mass, via
\begin{equation}
\label{eq:radius}
\langle r^2 \rangle_\Gamma = \frac{6}{M^2_\Gamma}\,.  
\end{equation}

Of particular importance is the charge radius of the vector form factor. This has been extracted from $\pi - e$ scattering data~\cite{Dally:1977vt,Dally:1981ur,Dally:1982zk,Amendolia:1984nz,Amendolia:1986wj,GoughEschrich:2001ji}, as well as $e^+ e^- \to \pi^+\pi^-$ data~\cite{Amendolia:1983di,Barkov:1985ac,Ananthanarayan:2017efc,Colangelo:2018mtw}. The decay $\tau\rightarrow\pi\pi\nu$ offers another channel to extract information about the pion form factor~\cite{Belle:2008xpe}. Also, the scalar radius is related to $\pi\pi$-scattering amplitudes~\cite{Donoghue:1990xh,Gasser:1990bv,Moussallam:1999aq}. Data coming from pion electroproduction in the moderate~\cite{JeffersonLabFpi:2000nlc,JeffersonLabFpi:2007vir,JeffersonLabFpi-2:2006ysh,JeffersonLab:2008gyl,JeffersonLab:2008jve,Horn:2007ug} and large~\cite{Bebek:1974iz,Bebek:1977pe,Brown:1973wr,Brauel:1979zk,Ackermann:1977rp,Amendolia:1985bs,Palestini:1985zc} $Q^2$ regions can also be used to constraint the scalar radius. The scalar radius is also of phenomenological interest, as it enters the chiral expansion of the pion decay constant.

For the parametrization, we focus on the results from the two-state fits to ensure that excited-states are eliminated. We apply the fit of Eq.~(\ref{eq:fit}) in three data sets: results obtained using the rest frame (R), the boosted frame (B), as well as a combination of results from both frames (R$\&$B). The monopole fit is performed on the form factors calculated from the two-state fit. We test different ranges for the $Q^2$ interval, that is, up to 0.55 GeV$^2$ for the three cases, as well as up to 1 and 2.5 GeV$^2$ for  B and R$\&$B. 

\begin{figure}[h!]
    \centering
    \includegraphics[scale=0.28]{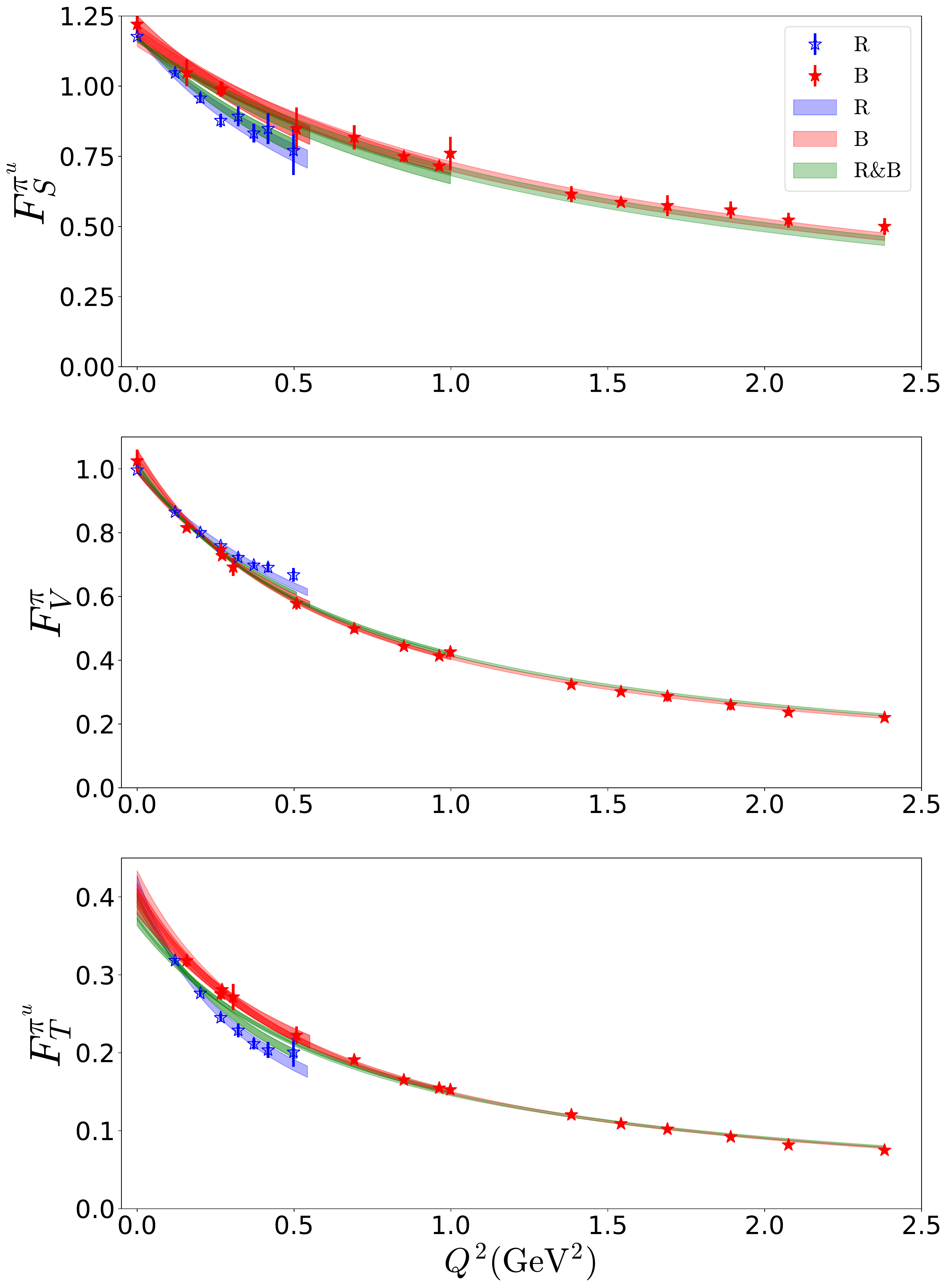}
    \caption{From top to bottom: The scalar, vector and tensor form factors of the pion using two-state fit in the rest (blue points) and boosted (red points) frame. The 2-parameter fitted form factors are shown with bands for the cases  R  (blue),  B (red), and  R$\&$B (green). The length of the band indicates the at $Q_{\rm max}^2$ interval used for the fit. }
    \label{fig:Pion_fit}
\end{figure}

In Table~\ref{tab:pion_fit} we give the fit parameters for each form factor. The one- and two-parameter fits are indicated with a subscript 1 and 2, respectively. Based on the results, there is a number of conclusions. First, the various estimates of $F_V(0)$ from any choice of $Q_{\rm max}$ and from 1- or 2-parameter fits are compatible. Also, the fitted values are compatible with actual lattice data. The same conclusions hold for $F_S(0)$. For the estimates of $\kappa_T$ ($=F_T(0)$) using the  B data sets at different $Q_{\rm max}$ are compatible. However, a slight tension is observed for the R$\&$B estimates as $Q_{\rm max}$ increases. The effect is also demonstrated in Fig.~\ref{fig:Pion_fit}, which will be discussed below. Conclusions can also be drawn for the monopole masses, where we find that the 1- and 2-parameter fits lead to compatible results. Slight tensions are observed as $Q_{\rm max}$ increases, as the determination of the monopole mass heavily relies on the slope, and therefore, the fit range. 

\begin{table}[h!]
\centering
\renewcommand{\arraystretch}{1.2}
\renewcommand{\tabcolsep}{6pt}
\begin{tabular}{l c c c c c c c c c}
 \\ [-3ex]
Frame & $Q^2_\mathrm{max}$ & $\,\,\,M^{\pi^u}_{S,1}\,\,\,$ &$\,\,\,F_{S,2}^{\pi^u}(0)\,\,\,$ &$\,\,\,M^{\pi^u}_{S,2}\,\,\,$ \hspace*{0.5cm}
      &$\,\,\,M^{\pi^u}_{V,1}\,\,\,$ &$\,\,\,F_{V,2}^{\pi^u}(0)\,\,\,$ &$\,\,\,M^{\pi^u}_{V,2}\,\,\,$  \hspace*{0.5cm}
      &$\,\,\,\kappa_T^{\pi^u}\,\,\,$ &$\,\,\,M^{\pi^u}_{T,2}\,\,\,$ \\[0.75ex]
    \hline
R & 0.55 & 0.959(54) & 1.177(5) & 0.958(56) & 0.927(19) & 0.991(6) & 0.938(22) & 0.414(14) & 0.633(39)   \\[0.5ex]
B & 0.55 & 1.066(74) & 1.217(34) & 1.075(76) & 0.831(50) & 1.011(25) & 0.853(38) & 0.394(17) & 0.808(62)  \\[0.5ex]
R$\&$B & 0.55 & 1.004(47) & 1.177(5) & 1.003(47) & 0.893(8) & 0.994(5) & 0.897(9) & 0.396(9) & 0.712(34)  \\[0.5ex]
\hline
B & 1.00 & 1.158(44) & 1.189(31) & 1.203(30) & 0.819(36) & 1.049(10) & 0.795(12) & 0.405(13) & 0.771(15)  \\[0.5ex]
R$\&$B  & 1.00 & 1.144(40) & 1.169(5) & 1.161(39) & 0.865(7) & 1.011(6) & 0.846(8) & 0.369(5) & 0.829(15)  \\[0.5ex]
\hline
B & 2.50 & 1.189(39) & 1.173(31) & 1.248(32) & 0.815(33) & 1.054(11) & 0.789(13) & 0.420(14) & 0.739(16)  \\[0.5ex]
R$\&$B & 2.50  & 1.201(36) & 1.165(6) & 1.221(36) & 0.855(7) & 1.017(6) & 0.832(8) & 0.376(5) & 0.800(12)  \\[0.5ex]
    \hline
\end{tabular}
\caption{Fit parameters for the up contribution to the pion as obtained from the two-state fits. The results in the rest, boosted and combined frames are indicated by R, B, and R$\&$B, respectively. The maximum momentum transfer, $Q^2_\mathrm{max}$, entering the fit is given in GeV$^2$, and the monopole masses in GeV. The subscript 1 and 2 indicate the one- and two-parameter fits, respectively. }
\label{tab:pion_fit}
\vspace*{0.2cm}
\end{table}

We choose as final results the values from the combined fit (R$\&$B) using a 2-parameter monopole Ansatz and the whole range of $Q^2$. 
\begin{eqnarray}
\label{eq:pion_fit_S}
F_{S}^{\pi^u}(0)    =  1.165(6)(4)\,,  &\hspace*{1cm}&
M^{\pi^u}_{S}    =  1.221(36)(60)\,{\rm GeV}\,, \\[1ex]
\label{eq:pion_fit_V}
F_{V}^{\pi^u}(0)    =  1.017(6)(6)\,, &\hspace*{1cm}&
M^{\pi^u}_{V}    =  0.832(8)(14)\,{\rm GeV}\,, \\[1ex]
\kappa_T^{\pi^u} =  0.376(5)(6)\,, &\hspace*{1cm}&
M^{\pi^u}_{T}    =  0.800(12)(29)\,{\rm GeV}\,.
\label{eq:pion_fit_T}
\end{eqnarray}
In the first parenthesis we give the statistical error and in the second parenthesis  the systematic error related to the fit range, determined as the difference between the values obtained using $Q^2=1$~GeV$^2$ and $Q^2_{\rm max}=2.5$ GeV$^2$.

In Fig.~\ref{fig:Pion_fit} we plot $F_{\cal O}(Q^2)$ using the two-state fit data in the rest and boosted frames. These are compared against the fitted form factors extracted from all cases of Table~\ref{tab:pion_fit}. For the scalar and vector case we show only the 2-parameter fits for better clarity. The main observation is that there is a small difference between the fits of the  R  data sets and the  B case obtained from $Q^2_{\rm max}=0.55$ GeV$^2$. As expected, the fits of R$\&$B for the same range are situated between the fits for the rest and boosted frame. For the fits on the combined data, we find compatibility for different values of $Q^2_{\rm max}$. The various fitted bands for $F_S$ and $F_V$ show that the corresponding value at $Q^2=0$ is in agreements. The discrepancy for the case of $\kappa_T$ discussed previously is due to the change in the slope for different data sets.  

The radii can be derived from the parametrization as discussed previously. Experimentally, the value of the pion charge radius has been extracted using data sets from pion electroproduction, pion electron scattering and positron electron annihilation into two charged pions. The averaged value is $\rr^{\pi^+}= 0.434(4)$ fm$^2$~\cite{Zyla:2020zbs}, which does not include the electroproduction data, due to uncertainties in the extraction of the radius.
Here, we apply Eq.~(\ref{eq:radius}) using the monopole mass from the one- and two-parameter fit. To test for systematic effects, we use the data from the rest and boosted frame separately, as well as the combined ones. For all cases we use the two-state fit data, and fit for all values of $Q^2$. The values for the radii can be found in Table~\ref{tab:pion_radii}. As expected, the tension observed in the monopole mass propagates to the radii. For each data set (rest, boosted, combined), we find agreement between the 1- and 2-parameter fits. However, given the difference in the slope there is tension between the three data sets. To address this point, we also give results by constraining the fit up to $Q^2=0.55$ GeV$^2$. 
Our final results for the radii are obtained from the 2-parameter monopole Ansatz applied on the R$\&$B data. Since the radii describe the small-$Q^2$ behavior, we choose $Q^2_{\rm max}=0.55$ GeV$^2$ and give as a systematic error the difference with the radius extracted from when fitting the  data up to 1 GeV$^2$
\begin{eqnarray}
\label{eq:pion_fit_rs}
\rr^{\pi^u}_S     &=&  0.232(22)(54)\,{\rm fm}^2\,,  \\[1ex]
\rr^{\pi^u}_V     &=&  0.291(6)(36)\,{\rm fm}^2\,, \\[1ex]
\rr^{\pi^u}_T     &=&  0.461(44)(121)\,{\rm fm}^2\,.
\label{eq:pion_fit_rT}
\end{eqnarray}
Our results for $\rr^{\pi^u}_S$ are compatible with the ones obtained in Ref.~\cite{Gulpers:2013uca} from the connected contributions on an $N_f=2$ ${\cal O}(a)$-improved Wilson fermions ensemble at a pion mass of 280 MeV. Similar values are also obtained from a 310 MeV pion mass ensemble of $N_f=2+1$ overlap fermions~\cite{Kaneko:2010ru}. We note that a sizeable logarithmic behavior in the pion mass is found in chiral perturbation theory~\cite{Gasser:1990bv,Bijnens:1998fm} which causes a rise in the radii. Therefore, at this stage, we do not attempt any comparison with the PDG value of $\rr^{\pi}_{V}$, as the ensemble we used is not at the physical value of the pion mass.

\begin{table}[h!]
\centering
\renewcommand{\arraystretch}{1.2}
\renewcommand{\tabcolsep}{6pt}
\begin{tabular}{l c c c c c c}
 \\ [-3ex]
Frame & $Q^2_{\rm max}$ & $\,\,\,\rr^{\pi^u}_{S,1}\,\,\,$  &$\,\,\,\rr^{\pi^u}_{S,2}\,\,\,$ \hspace*{0.5cm}
      &$\,\,\,\rr^{\pi^u}_{V,1}\,\,\,$  &$\,\,\,\rr^{\pi^u}_{V,2}\,\,\,$  \hspace*{0.5cm}
        &$\,\,\,\rr^{\pi^u}_{T,2}\,\,\,$ \\[0.75ex]
    \hline
R & 0.55 & 0.254(29) & 0.254(30) & 0.272(11) & 0.265(13) & 0.582(72)  \\[0.5ex]
B & 0.55 & 0.206(29) & 0.202(29) & 0.339(41) & 0.321(29) & 0.357(55) \\[0.5ex]
R$\&$B & 0.55 & 0.232(22) & 0.232(22) & 0.293(6) & 0.291(6) & 0.461(44) \\[0.5ex]
\hline
B & 1.00 & 0.174(13) & 0.174(13) & 0.348(31) & 0.370(11) & 0.393(16) \\[0.5ex]
R$\&$B & 1.00 & 0.178(12) & 0.173(12) & 0.312(5) & 0.327(6) &  0.340(12) \\[0.5ex]
\hline
B & 2.50 & 0.165(11)  & 0.150(8) & 0.352(28) & 0.375(12) & 0.427(18)   \\[0.5ex]
R$\&$B & 2.50 & 0.162(10) & 0.157(9) & 0.320(5) & 0.337(7) & 0.365(11)  \\[0.5ex]
    \hline
\end{tabular}
\caption{The scalar, charge and tensor radii for the pion in fm$^2$. The notation is the same as Table~\ref{tab:pion_fit}.}
\label{tab:pion_radii}
\vspace*{0.2cm}
\end{table}

As already mentioned, $F_T(0)$ can be obtained from fits applied on the tensor form factor to parametrize its $Q^2$ dependence. It has been discussed, for the nucleon and pion case, that the $Q^2$ dependence of the vector and tensor form factors are expected to be the same due to the elastic unitarity relation~\cite{Hoferichter:2018zwu}. This argument holds for $Q^2$ below 1 GeV$^2$ due to corrections from inelastic states beyond that. We test this argument for the pion by taking the ratio of the tensor and vector form factors. If the argument holds, the ratio should be constant with $Q^2$ and equal to $F_T(0)$.

\begin{figure}[h!]
    \centering
    \includegraphics[scale=0.28]{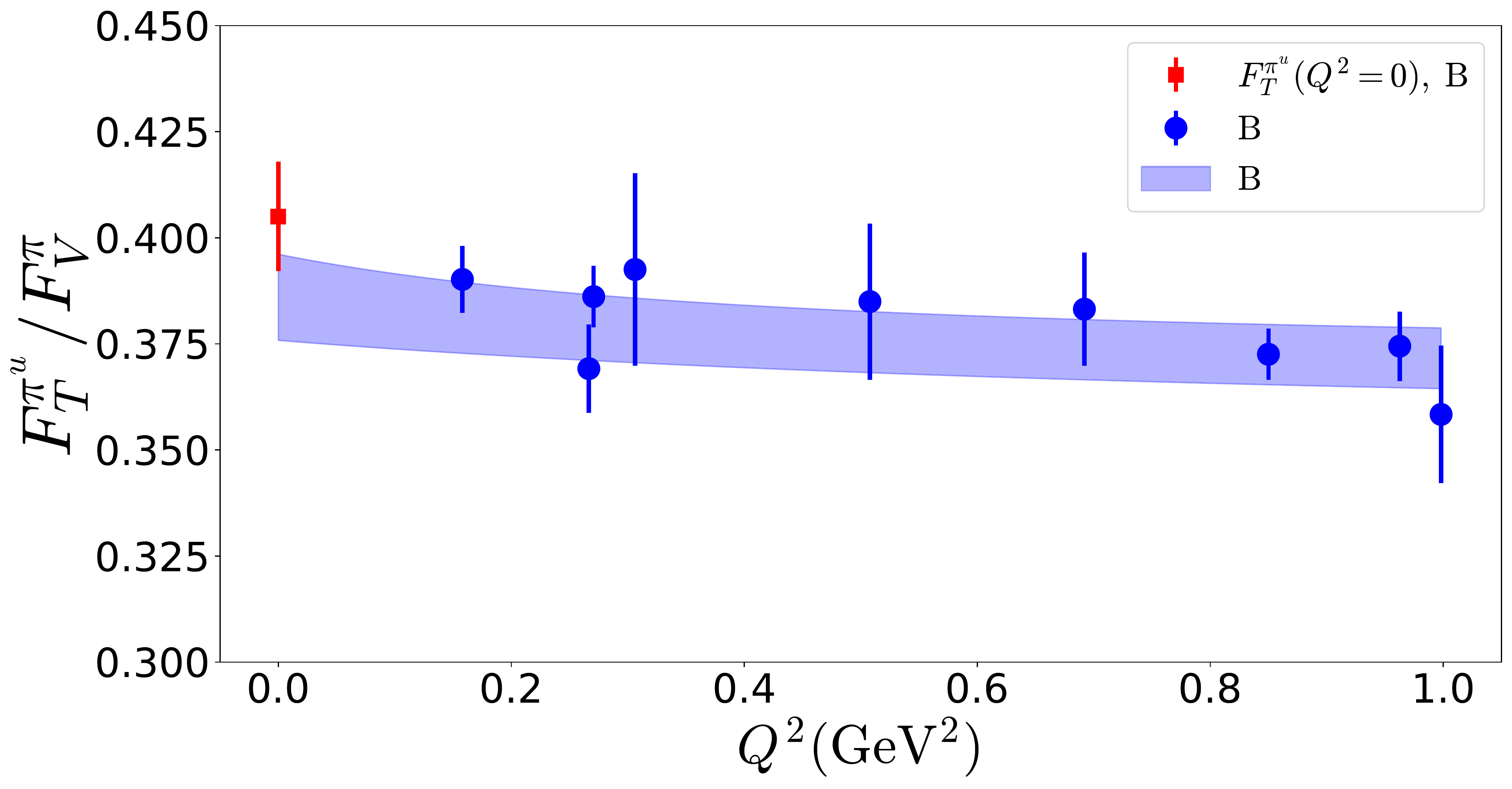}
    \caption{The ratio $F_T/F_V$ as a function of $Q^2$ using the two-state fit results from the boosted frame (blue circles). The curve is calculated from the 2-parameter monopole fit applied on all available data for the vector and tensor form factors. The red square is $F_T(0)$ calculated from the 2-parameter fit using all available lattice data (R$\&$B). }
    \label{fig:T_over_V_FFs_pion}
\end{figure}
The ratio is presented in Fig.~\ref{fig:T_over_V_FFs_pion}, using lattice data and their parametrizations. The data point at $Q^2=0$ is obtained from the actual lattice data of $F_V(0)$ and the 2-parameter fit using all available lattice data for $F_T(0)$ up to $Q=1 \,{\rm GeV}^2$. The ratio using the monopole fit for both form factors is also shown. It is interesting to observe that the $Q^2$ behavior of the ratio is very mild compared to the individual form factors, suggesting that it is largely canceled between the two form factors. The value of the vector form factor at $Q^2=2.5$ GeV$^2$ is 20$\%$ its value at $Q^2=0$. Similar comparison for the tensor form factor shows a similar change, that is, 75$\%$. For the ratio of the two form factors, we find that the change is only about 10-15$\%$. A reliable extraction of $F_T(0)$ is desirable, as it enters the average transverse shift of transversely polarized quarks (see Sec.~\ref{sec:impact_b}), and it is the tensor anomalous magnetic moment. Estimations from both the monopole fit and the ratio of the tensor and vector form factors are expected to be in better agreement once various sources of systematic uncertainties are better controlled.

\vspace*{1.25cm}
\section{Kaon Form Factors}
\label{sec_kaon}

\subsection{Excited-states investigation}

\begin{figure}[h!]
    \centering
    \includegraphics[scale=0.28]{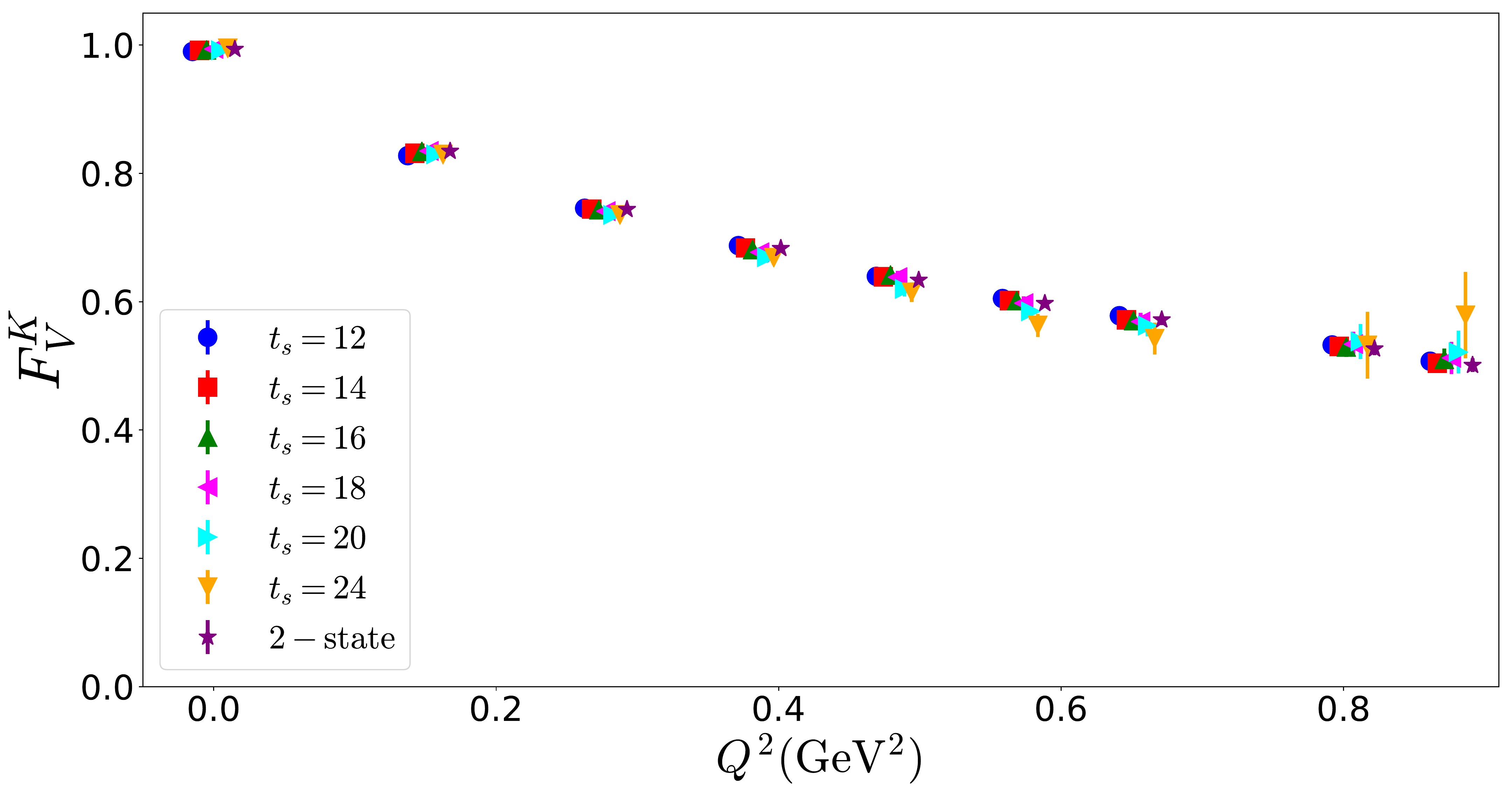}
    \caption{The vector form factor for the kaon in the rest frame. The notation is the same as Fig.~\ref{fig:FFs_pion_rest}.}
    \label{fig:FK_v_rest}
\end{figure}

In this section, we apply the analysis method discussed above for the up- and strange-quark contributions to the kaon form factors. We begin our presentation with the results in the rest frame shown in Fig.~\ref{fig:FK_v_rest}. We can obtain the form factors up to $Q^2\sim 1$ GeV$^2$, as compared to about 0.5 GeV$^2$ for the pion. This is expected, as the kaon is about twice as heavy as the pion for this ensemble. In Fig.~\ref{fig:FK_v_rest} we show the vector form factor after combining the up- and strange-quark contributions, each multiplied by their respective charge. Similar to the pion case, we find that excited-states contamination are suppressed, as the results for the various $t_s$ values are compatible. Also, agreement is found between the two-state fits and the plateau values. The decay of the form factor with $Q^2$ is slower than in the case of the pion. This is an indication of SU(3) flavor symmetry breaking effect. We will revisit this discussion in Sec.~\ref{sec:SU3}.
\vspace*{0.5cm}

\begin{figure}[h!]
    \centering
    \includegraphics[scale=0.28]{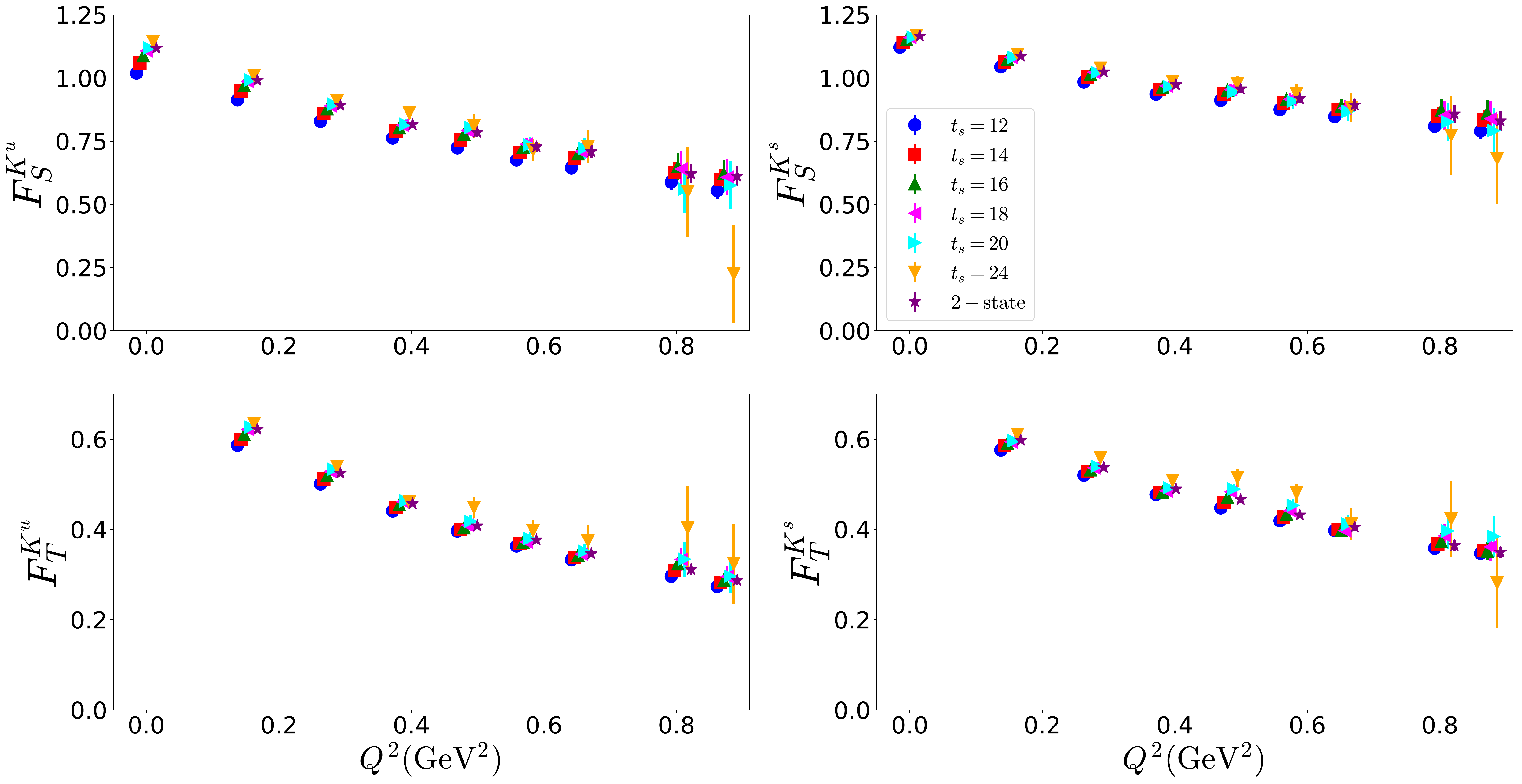}
\vskip -.25cm
    \caption{The scalar (top panel) and tensor (bottom panel) form factor for the up (left) and strange (left) contributions to the kaon. The data have been obtained in the rest frame. The notation is the same as Fig.~\ref{fig:FFs_pion_rest}.}
    \label{fig:FK_s_t_rest}
\end{figure}

The rest-frame results for the scalar and tensor form factors for the individual flavors are shown in Fig.~\ref{fig:FK_s_t_rest}. As for the pion, the up-quark contributions are susceptible to excited-states contamination; a convergence to the ground state is observed at $t_s/a=18$ for the scalar form factor, and $t_s/a=16$ for the tensor one. The plateau values at this and larger time separations are compatible with the two-state fit. 
For the strange-quark form factors we find smaller contamination from excited states. For both flavors, we choose the two-state fits as final results.

\begin{figure}[h!]
    \centering
    \includegraphics[scale=0.28]{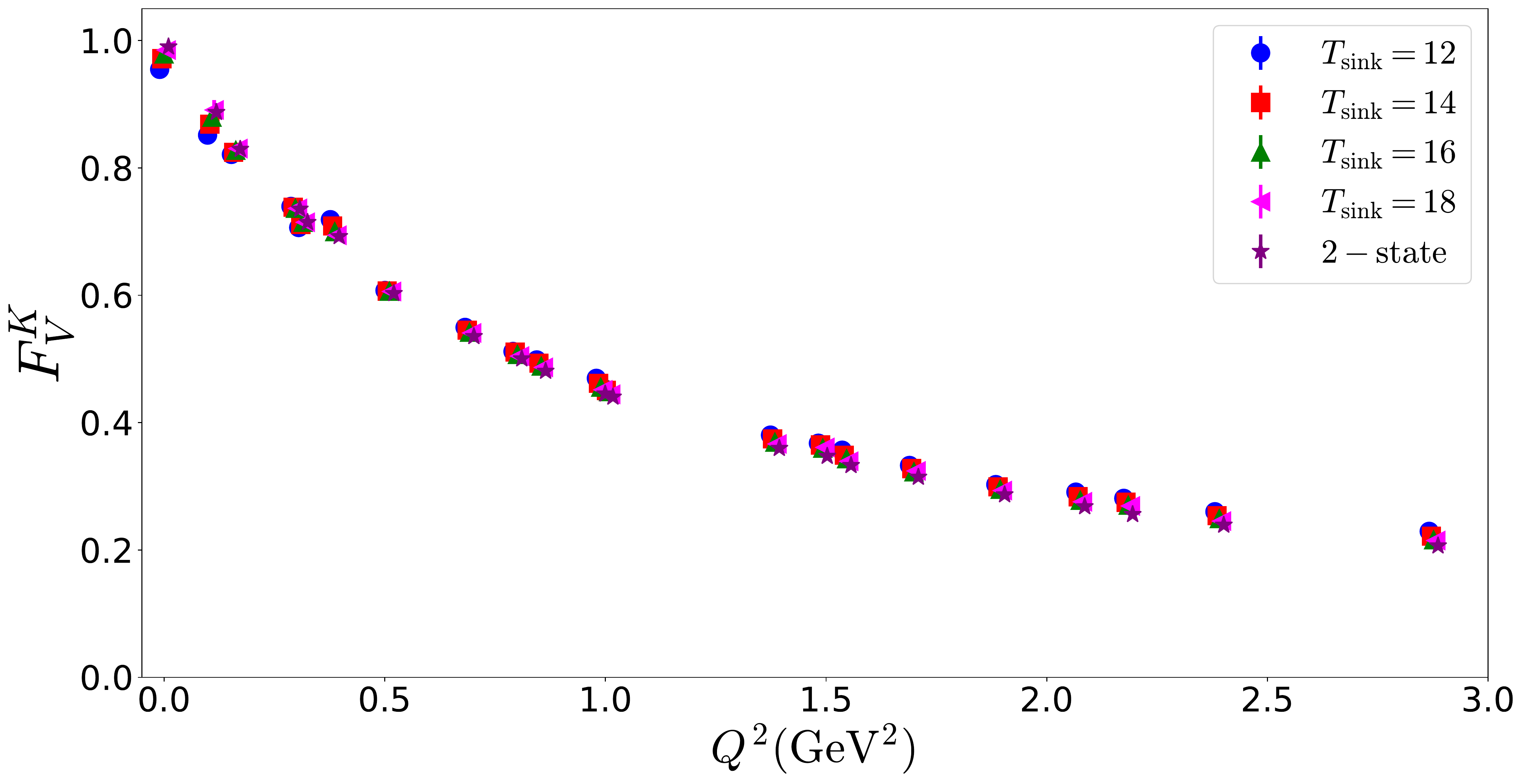}
    \caption{The vector form factor for the kaon in the boosted frame. The notation is the same as Fig.~\ref{fig:FFs_pion_boost}. }
    \label{fig:FK_v_boost}
\end{figure}

\newpage
The vector form factor in the boosted frame is shown in Fig.~\ref{fig:FK_v_boost} for the four values of the source-sink time separation and the two-state fit. We see that there are mild excited states effects in the small-$Q^2$ region, which are suppressed at $t_s=18a$. The removal of the excited-states contamination changes the slope of the form factor at small $Q^2$ values making it more steep. Also the form factor at $Q^2=0$ becomes equal to one in the two-state fit, as expected from charge conservation.

\begin{figure}[h!]
    \centering
     \includegraphics[scale=0.28]{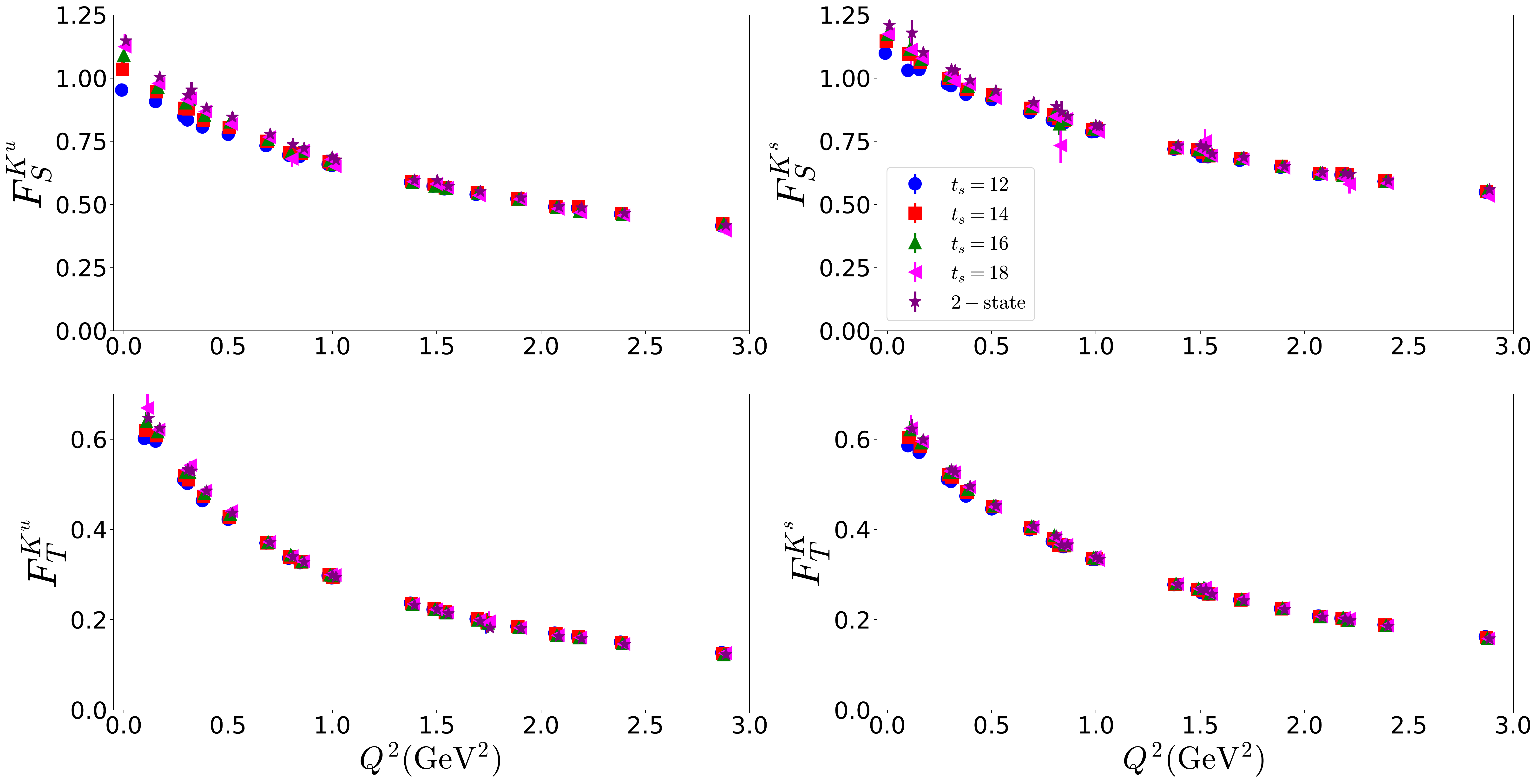}
    \caption{The scalar and tensor form factors in the boosted frame. The notation is the same as Fig.~\ref{fig:FFs_pion_boost}. }
    \label{fig:FK_s_t_boost}
\end{figure}

The scalar and tensor form factors are shown in Fig.~\ref{fig:FK_s_t_boost}. The up-quark part of the tensor form factor, $F_T^{K^u}$, exhibits excited-states contamination for $Q^2$ up to 0.5 GeV$^2$, which are, however, eliminated for $t_s \ge 16a$. For the scalar case, the excited-states effect extend up to $Q^2=0.8$ GeV$^2$, but are suppressed for $t_s=18a$. The strange-quark contribution for both form factors is less affected by excited states. It is also worth noting that the tensor form factor for the kaon is about a factor of two larger than for the pion. This is expected because the suppression of $F_T$ by the mass, indicating that the ratio $F^K_T/F^\pi_T$ is about $M_K/M_\pi\sim 2$.

\begin{figure}[h!]
    \centering
    \includegraphics[scale=0.27]{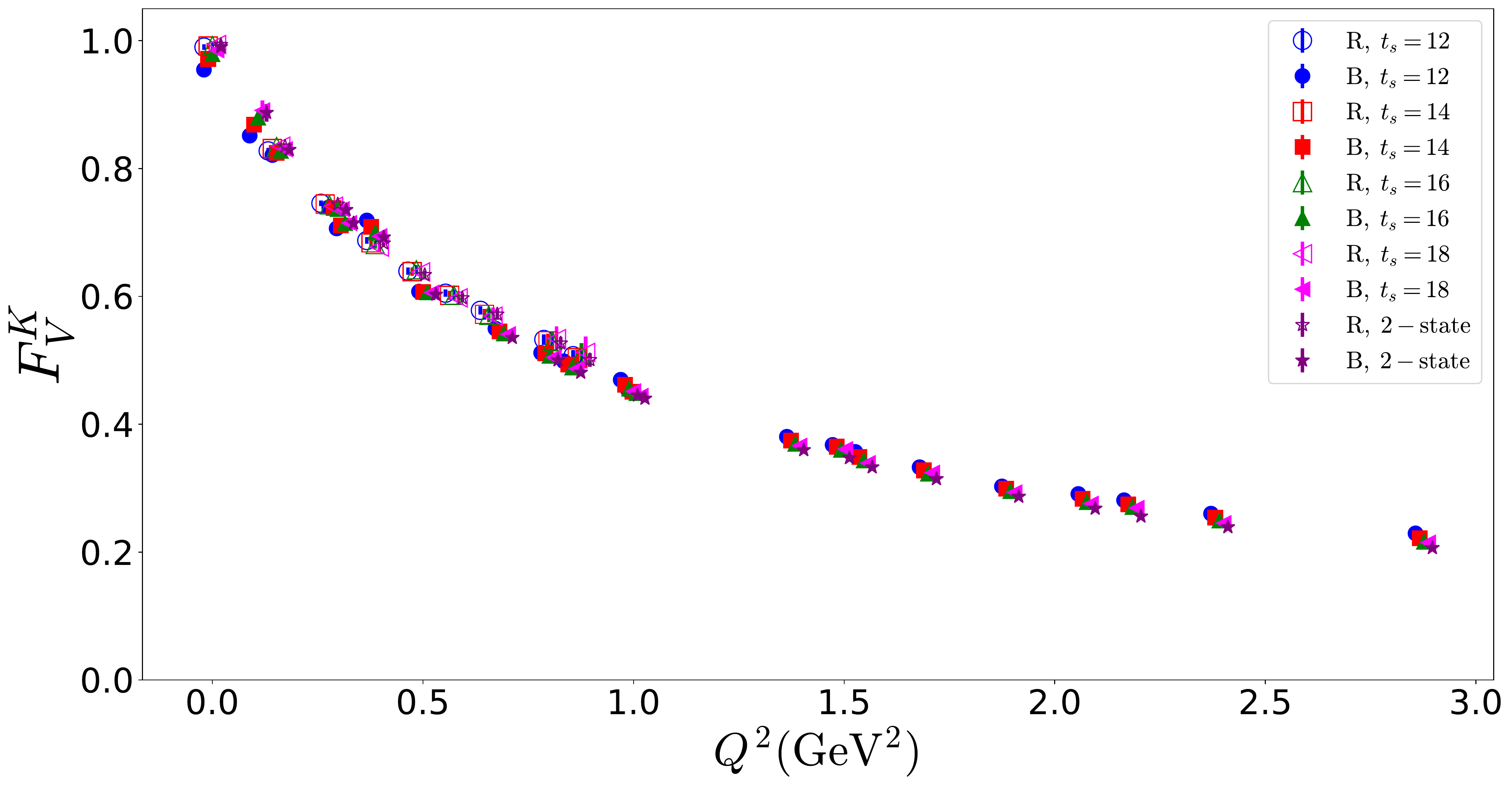}
    \caption{Comparison of the vector form factor of the kaon between the rest and boosted frame. The notation is the same as Fig.~\ref{fig:FFs_pion_boost}. }
    \label{fig:FK_v_all}
\end{figure}

\begin{figure}[h!]
    \centering
     \includegraphics[scale=0.25]{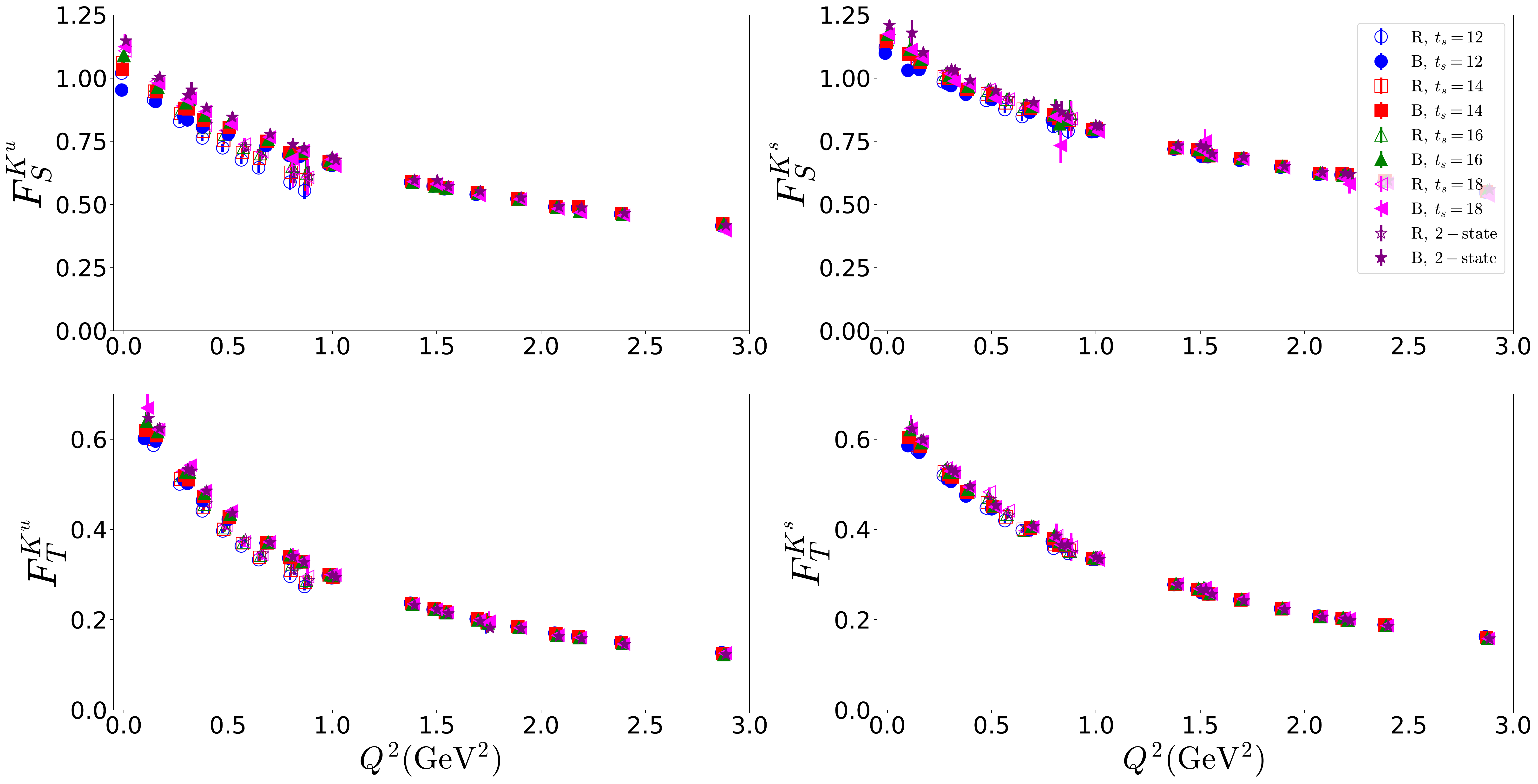}
    \caption{Comparison of the scalar and tensor form factor of the kaon between the rest and boosted frame. The notation is the same as Fig.~\ref{fig:FFs_pion_boost}. }
    \label{fig:FK_s_t_all}
\end{figure}

\vspace*{0.5cm}
The comparison between the rest and boosted frame is shown in Fig.~\ref{fig:FK_v_all} for the vector and Fig.~\ref{fig:FK_s_t_all} for the scalar and tensor. The results for the vector are fully compatible for the $t_s$ values that excited states effects are eliminated. This is expected because excited states are frame dependent. 
The strange-quark scalar and tensor form factors show compatibility between the two frames. The situation with the up-quark form factors is similar with the pion: there is some tension in the slope between the two frames, with the rest-frame results having a steeper slope.

\subsection{Parametrization of kaon form factors}
\label{sec:fits_kaon}

Using the setup outlined in Sec.~\ref{sec:fits_pion}, we parametrize the $Q^2$ dependence of the kaon form factors using the monopole fit of Eq.~\eqref{eq:fit}. Similar to the pion case, we explore both a 1- and 2-parameter fits. Here we employ two values for the $Q^2_{\rm max}$, that is 1 and 3 GeV$^2$. First we explore the parametrization of the vector form factor, which combines the up- and strange-quark contributions, shown in Table~\ref{tab:kaon_vector_fit}. We find that $M_V$ is insensitive to the choice of fit (1- or 2-parameters). We also observe that the fitted $F_V(0)$ is fully compatible to the value obtained directly from the matrix elements. The results for different fit ranges and including data sets from different frames are also compatible.

\begin{table}[h!]
\centering
\renewcommand{\arraystretch}{1.2}
\renewcommand{\tabcolsep}{6pt}
\begin{tabular}{l c c c c c c c c c}
 \\ [-3ex]
Frame & $Q^2_\mathrm{max}$      &$\,\,\,M^{K}_{V,1}\,\,\,$ &$\,\,\,F_{V,2}^{K}\,\,\,$ &$\,\,\,M^{K}_{V,2}\,\,\,$   \\[0.75ex]
    \hline
R & 1.0 & 0.927(11) & 0.987(5) & 0.937(13)  \\[0.5ex]
B & 1.0 & 0.913(8) & 1.015(5) & 0.885(5)    \\[0.5ex]
R$\&$B & 1.0 & 0.911(6) & 1.004(4) & 0.898(4)   \\[0.5ex]
\hline
B & 3.0 & 0.901(7) & 1.028(5) & 0.867(5)  \\[0.5ex]
R$\&$B & 3.0 & 0.899(6) & 1.015(4) & 0.879(5)   \\[0.5ex]
\hline
\end{tabular}
\caption{Fit parameters for the kaon vector form factor as obtained from the two-state fits. The results in the rest, boosted and combined frames are indicated by R, B, and R$\&$B, respectively. The maximum momentum transfer, $Q^2_\mathrm{max}$, entering the fit is given in GeV$^2$, and the monopole masses in GeV. The subscript 1 and 2 indicate the one- and two-parameter fits, respectively.}
\label{tab:kaon_vector_fit}
\vspace*{0.2cm}
\end{table}

\begin{figure}[h!]
    \centering
    \includegraphics[scale=0.24]{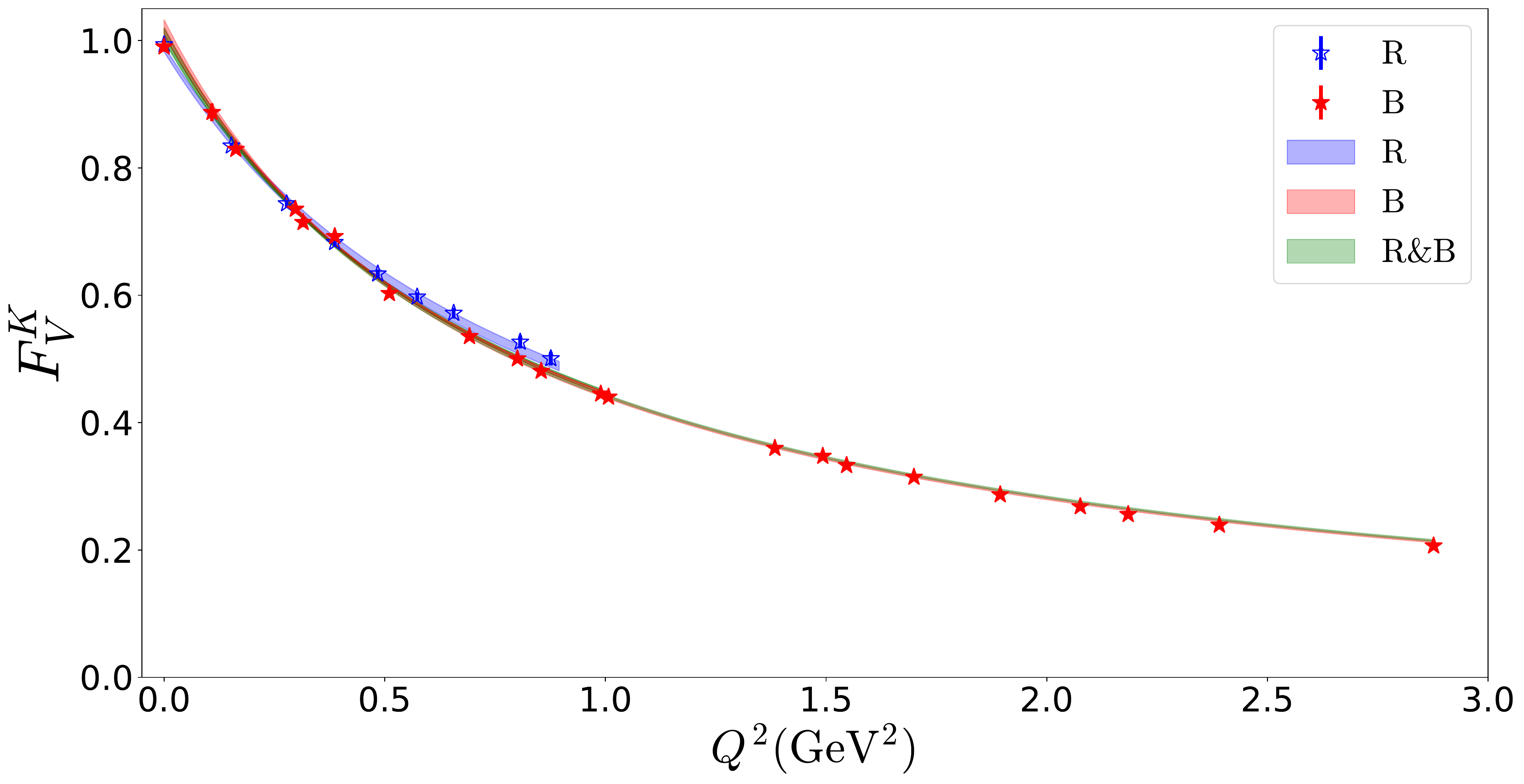}
    \caption{Parametrization of the vector form factor for the kaon. The notation is the same as Fig.~\ref{fig:Pion_fit}.}
    \label{fig:kaon_fit_v}
\end{figure}

\begin{figure}[h!]
    \centering
    \includegraphics[scale=0.28]{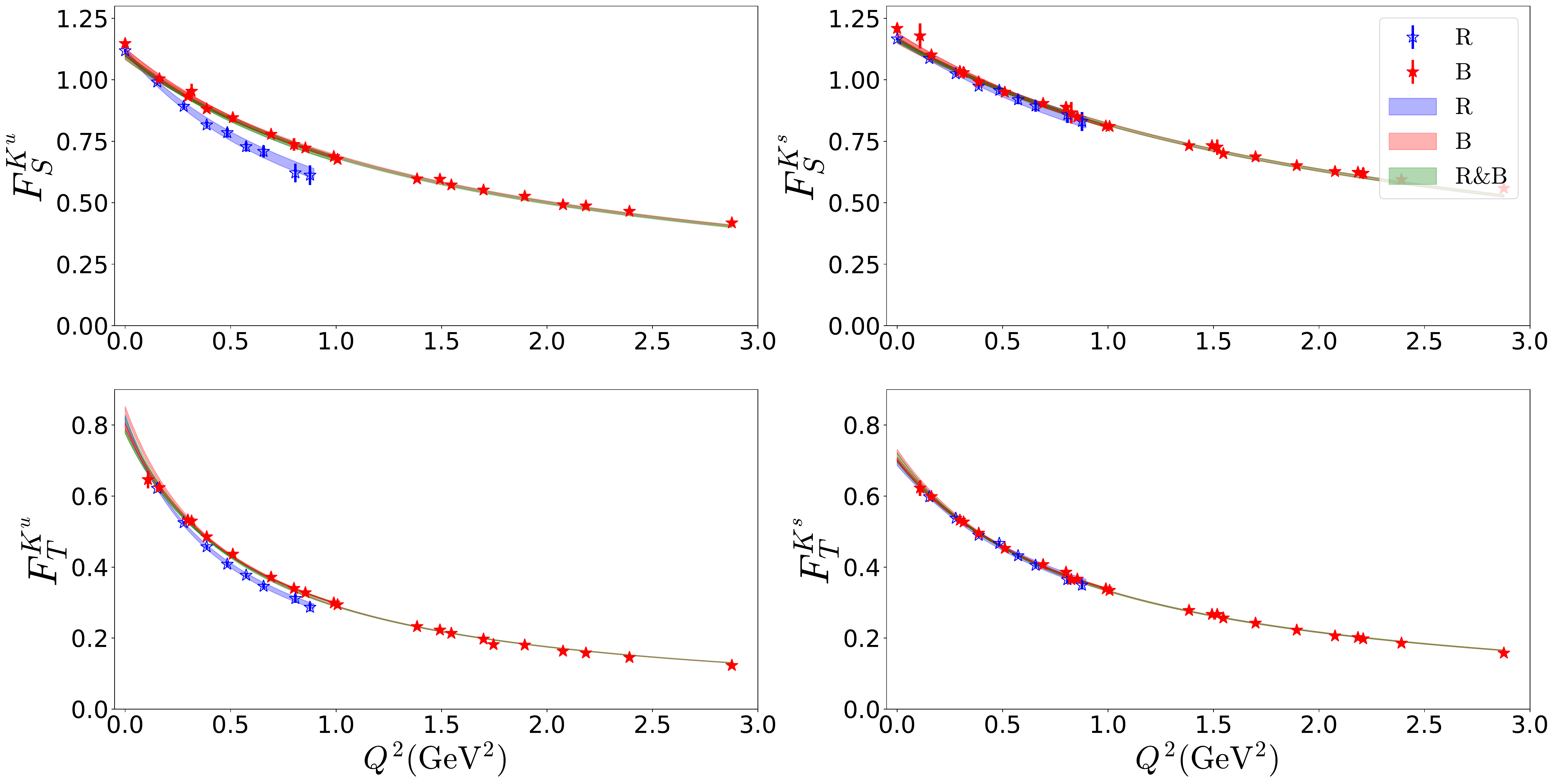}
    \caption{Parametrization of the scalar and tensor form factor for the up (left) and strange (right) quark components of the kaon. The notation is the same as Fig.~\ref{fig:Pion_fit}.}
    \label{fig:kaon_fit_s_t}
\end{figure}

\bigskip
In Tables~\ref{tab:kaon_fit_u} - \ref{tab:kaon_fit_s} we provide the parameters of the fits on all form factors for the up- and strange-quark components, respectively. In summary, we find no dependence on the number of parameters, and the fitted $F_{\cal O}(0)$ is independent of the fit range and the data sets included. Moreover, there is agreement with the actual lattice data for the scalar and vector cases. Compatible results are also seen across all estimates of the monopole mass for the strange quark contributions. However, the estimates of $M^{K^u}_S$ extracted from the rest frame has tension with the value obtained from the boosted frame and the combined case. Similar behavior is also observed in $M^{K^u}_T$. All the aforementioned conclusions can be seen in Figs.~\ref{fig:kaon_fit_v} - \ref{fig:kaon_fit_s_t}.

\newpage

\begin{table}[h!]
\centering
\renewcommand{\arraystretch}{1.2}
\renewcommand{\tabcolsep}{6pt}
\begin{tabular}{l l c c c c c c c c}
 \\ [-3ex]
Frame & $Q^2_\mathrm{max}$ &$\,\,\,M^{K^u}_{S,1}\,\,\,$ &$\,\,\,F_{S,2}^{K^u}(0)\,\,\,$ &$\,\,\,M^{K^u}_{S,2}\,\,\,$ \hspace*{0.5cm}
      &$\,\,\,M^{K^u}_{V,1}\,\,\,$ &$\,\,\,F_{V,2}^{K^u}(0)\,\,\,$ &$\,\,\,M^{K^u}_{V,2}\,\,\,$  \hspace*{0.5cm}
      &$\,\,\,\kappa_T^{K^u}\,\,\,$ &$\,\,\,M^{K^u}_{T,2}\,\,\,$ \\[0.75ex]
    \hline
R & 1.0 & 1.054(35) & 1.119(7) & 1.052(38) & 0.857(10) & 0.989(5) & 0.862(12)  & 0.816(13) & 0.701(20) \\[0.5ex]
B & 1.0 &  1.201(24) & 1.113(15) & 1.258(15) & 0.862(8) & 1.019(5) & 0.828(5) & 0.799(8) & 0.772(4)  \\[0.5ex]
R$\&$B & 1.0 & 1.223(18) & 1.103(8) & 1.251(14) & 0.853(5) & 1.005(4) & 0.841(4) & 0.783(6) & 0.782(4)  \\[0.5ex]
\hline
B & 3.0 & 1.233(20) & 1.096(14) & 1.301(15) & 0.848(7) & 1.036(6) & 0.807(5) & 0.843(9) & 0.724(5)   \\[0.5ex]
R$\&$B & 3.0 & 1.255(16) & 1.093(8) & 1.291(15) & 0.841(5) & 1.016(5) & 0.822(5) & 0.844(9) & 0.724(5)  \\[0.5ex]
    \hline
\end{tabular}
\caption{Fit parameters for the up-quark contribution to the kaon as obtained from the two-state fits. The results in the rest, boosted and combined frames are indicated by R, B, and R$\&$B, respectively. The maximum momentum transfer, $Q^2_\mathrm{max}$, entering the fit is given in GeV$^2$, and the monopole masses in GeV. The subscript 1 and 2 indicate the one- and two-parameter fits, respectively. }
\label{tab:kaon_fit_u}
\vspace*{0.2cm}
\end{table}

\begin{table}[h!]
\centering
\renewcommand{\arraystretch}{1.2}
\renewcommand{\tabcolsep}{6pt}
\begin{tabular}{l l c c c c c c c c}
 \\ [-3ex]
Frame & $Q^2_\mathrm{max}$ &$\,\,\,M^{K^s}_{S,1}\,\,\,$ &$\,\,\,F_{S,2}^{K^s}(0)\,\,\,$ &$\,\,\,M^{K^s}_{S,2}\,\,\,$ \hspace*{0.5cm}
      &$\,\,\,M^{K^s}_{V,1}\,\,\,$ &$\,\,\,F_{V,2}^{K^s}(0)\,\,\,$ &$\,\,\,M^{K^s}_{V,2}\,\,\,$  \hspace*{0.5cm}
      &$\,\,\,\kappa_T^{K^s}\,\,\,$ &$\,\,\,M^{K^s}_{T,2}\,\,\,$ \\[0.75ex]
    \hline
R & 1.0 & 1.471(60) & 1.166(6) & 1.486(71) & 1.100(15) & 0.986(5) & 1.116(19) & 0.694(8) & 0.972(33)  \\[0.5ex]
B & 1.0 & 1.409(31) & 1.179(12) & 1.479(16) & 1.025(10) & 1.014(4) & 1.007(5) & 0.704(6) & 0.961(5)  \\[0.5ex]
R$\&$B & 1.0 & 1.507(22) & 1.166(7) & 1.506(16) & 1.041(8) & 1.006(3) & 1.022(4) & 0.700(4) & 0.967(5)  \\[0.5ex]
\hline
B & 3.0 & 1.462(25) & 1.158(11) & 1.553(17) & 1.015(8) & 1.024(5) & 0.989(6) & 0.725(6) & 0.921(6)  \\[0.5ex]
R$\&$B & 3.0 & 1.536(19) & 1.158(7) & 1.552(17) & 1.028(7) & 1.017(4) & 1.000(6) & 0.717(5) & 0.930(6)  \\[0.5ex]
    \hline
\end{tabular}
\caption{Fit parameters for the strange-quark contribution to the kaon as obtained from the two-state fits. The results in the rest, boosted and combined frames are indicated by R, B, and R$\&$B, respectively. The maximum momentum transfer, $Q^2_\mathrm{max}$, entering the fit is given in GeV$^2$, and the monopole masses in GeV. The subscript 1 and 2 indicate the one- and two-parameter fits, respectively.}
\label{tab:kaon_fit_s}
\vspace*{0.2cm}
\end{table}

Similarly to the pion, we choose as final results the values from the combined fit (R$\&$B) using a 2-parameter monopole ansatz and $Q^2_{\rm max}=3$ GeV$^2$, which leads to
\begin{eqnarray}
\label{eq:kaon_fit_S}
F_{V}^{K}(0)    =  1.015(4)(11)\,, &\hspace*{1cm}&
M^{K}_{V}    =  0.879(5)(19)\,{\rm GeV}\,,
\end{eqnarray}
for the vector case. For the up-quark contributions to the scalar and tensor parameters we obtain
\begin{eqnarray}
F_{S}^{K^u}(0)    =  1.093(8)(10)\,,  &\hspace*{1cm}&
M^{K^u}_{S}    =  1.291(15)(40)\,{\rm GeV}\,, \\[1ex]
\label{eq:kaon_fit_V_u}
F_{V}^{K^u}(0)    =  1.016(5)(11)\,,  &\hspace*{1cm}&
M^{K^u}_{V}    =  0.822(5)(19)\,{\rm GeV}\,, \\[1ex]
\kappa_T^{K^u} =  0.844(9)(61)\,, &\hspace*{1cm}&
M^{K^u}_{T}    =  0.724(5)(59)\,{\rm GeV}\,.
\label{eq:kaon_fit_T_u}
\end{eqnarray}
The corresponding values for the strange-quark components are 
\begin{eqnarray}
F_{S}^{K^s}(0)    =  1.158(7)(8)\,,  &\hspace*{1cm}&
M^{K^s}_{S}    =  1.552(17)(46)\,{\rm GeV}\,, \\[1ex]
\label{eq:kaon_fit_V_s}
F_{V}^{K^s}(0)    =  1.017(4)(11)\,,  &\hspace*{1cm}&
M^{K^s}_{V}    =  1.000(6)(22)\,{\rm GeV}\,, \\[1ex]
\kappa_T^{K^s} =  0.717(5)(17)\,, &\hspace*{1cm}&
M^{K^s}_{T}    =  0.930(6)(37)\,{\rm GeV}\,.
\label{eq:kaon_fit_T_s}
\end{eqnarray}
Overall, we find that the systematic uncertainties dominate the statistical ones. In particular, the extraction of the monopole masses is more susceptible to the fit range for $Q^2$, which is reflected in the systematic errors indicated in the second parenthesis. This error corresponds to the difference in the estimate between $Q^2_{\rm max}=1$ GeV$^2$ and $Q^2_{\rm max}=3$ GeV$^2$. 

\vspace*{1cm}

\begin{table}[h!]
\centering
\renewcommand{\arraystretch}{1.2}
\renewcommand{\tabcolsep}{6pt}
\begin{tabular}{l l c c c c c c c c}
 \\ [-3ex]
Frame & $Q^2_\mathrm{max}$ &$\,\,\,\rr^{K^u}_{S,1}\,\,\,$ &$\,\,\,\rr_{S,2}^{K^u}\,\,\,$ &$\,\,\,\rr^{K^s}_{S,1}\,\,\,$ &$\,\,\,\rr^{K^s}_{S,2}\,\,\,$ \hspace*{0.5cm}
      &$\,\,\,\rr^{K}_{V,1}\,\,\,$  &$\,\,\,\rr^{K}_{V,2}\,\,\,$  \hspace*{0.5cm}
        &$\,\,\,\rr^{K^u}_{T,2}\,\,\,$  \hspace*{0.15cm}
        &$\,\,\,\rr^{K^s}_{T,2}\,\,\,$ \\[0.75ex]
    \hline
R & 1.0 & 0.210(14) & 0.211(15) & 0.108(9) & 0.106(10) & 0.272(6) & 0.266(8) & 0.475(27) & 0.247(17)  \\[0.5ex]
B & 1.0 & 0.162(6) & 0.148(4) & 0.118(5) & 0.107(2) & 0.280(5) & 0.298(3) & 0.392(5) & 0.253(3)   \\[0.5ex]
R$\&$B & 1.0 & 0.156(5) & 0.149(3) & 0.103(3) & 0.103(2) & 0.282(4) & 0.289(3) & 0.382(4) & 0.250(3)  \\[0.5ex]
\hline
B & 3.0 & 0.154(5) & 0.138(3) & 0.109(4) & 0.097(2) & 0.288(5) & 0.311(4) & 0.445(6) & 0.275(3)   \\[0.5ex]
R$\&$B & 3.0 & 0.147(4) & 0.139(3) & 0.099(2) & 0.097(2) & 0.289(4) & 0.302(3) & 0.427(5) & 0.270(3)  \\[0.5ex]
    \hline
\end{tabular}
\caption{The scalar, charge and tensor radii of the kaon form factors in fm$^2$.}
\label{tab:kaon_radii}
\vspace*{0.2cm}
\end{table}

The behavior of the monopole mass with respect to the fit range is also observed in the radii. More precisely, the values of $\rr_{S}^{K^u}$ and  $\rr_{T}^{K^u}$  obtained from the rest frame data are higher then the corresponding one in the boosted-frame and combined-frame fits. On the contrary, we find full compatibility for the case of the strange-quark contributions. Experimentally, the charge radius of the kaon has been measured from the $K^-$ form factor in $K e$ elastic scattering~\cite{Dally:1980dj,Amendolia:1986ui} and the PDG value reported is $\rr^K= 0.314(25)$~\cite{Zyla:2020zbs}. Additional data for the kaon form factor has been obtained at moderate and large $Q^2$ using electroproduction processes~\cite{Brauel:1979zk,Carmignotto:2018uqj}.

We report the following as final values from this analysis
\begin{eqnarray}
\rr^{K^u}_S     =  0.149(3)(10)\,{\rm fm}^2\,,  &\hspace*{1cm}& \rr^{K^s}_S     =  0.103(2)(6)\,{\rm fm}^2\,,  \\[1ex]
\rr^{K}_V     =  0.289(3)(13)\,{\rm fm}^2\,, &\hspace*{1cm}& \\[1ex]
\rr^{K^u}_T     =  0.382(4)(45)\,{\rm fm}^2\,, &\hspace*{1cm}& \rr^{K^s}_T     =  0.250(3)(20)\,{\rm fm}^2\,,
\label{eq:kaon_fit_rT}
\end{eqnarray}
based on the same criteria as for the pion. We observe that the extraction of the tensor radius is more sensitive to the fit range. We note that our results for $\rr^{K}_V$ are compatible with the ones of Ref.~\cite{Kaneko:2010ru} obtained from an $N_f=2+1$ ensemble of overlap fermions producing a pion mass of 310 MeV.

\begin{figure}[h!]
    \centering
    \includegraphics[scale=0.27]{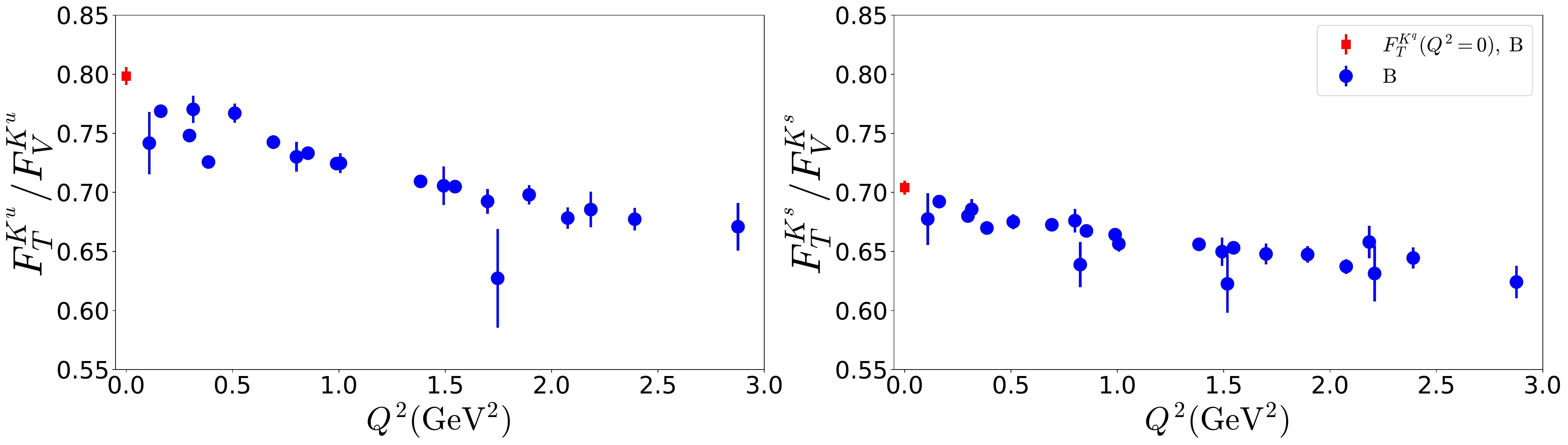}
\vspace*{-0.4cm}
    \caption{The ratio of up- and strange-quark contributions to the kaon tensor over the vector form factor (red squares) and over the corresponding quark contributions to the vector form factor (green triangles). The notations is the same as Fig.~\ref{fig:T_over_V_FFs_pion}. }
    \label{fig:T_over_V_FFs_kaon}
\end{figure}

The argument of Ref.~\cite{Hoferichter:2018zwu} that the $Q^2$-dependence of the vector and tensor form factors should match, could be also be discussed for the kaon. One can look at the isovector combination from dispersive
arguments and the isoscalar one using Vector Meson Dominance. This implicates the $\rho,\,\omega$, and $\phi$ mesons~\cite{Martin_private}. In Fig.~\ref{fig:T_over_V_FFs_kaon} we show the ratio of the tensor and vector form factors for each quark flavor. Interestingly, we find that the ratios have smaller slopes compared to the individual form factors. Nevertheless, a slope is still observed. For the strange quark, the ratio changes by about $10-15\%$. On the other hand, for the up quark we find an effect of $\sim15-20\%$. This is in contrast to the pion, where the effect is less than $\sim5\%$. This sizeable slope is expected to be due to SU(3) flavor symmetry breaking effect, which is found to be about $20\%$ (see Sec.~\ref{sec:SU3}).

\section{SU(3) flavor symmetry breaking}
\label{sec:SU3}

\begin{figure}[h!]
    \centering
    \includegraphics[scale=0.24]{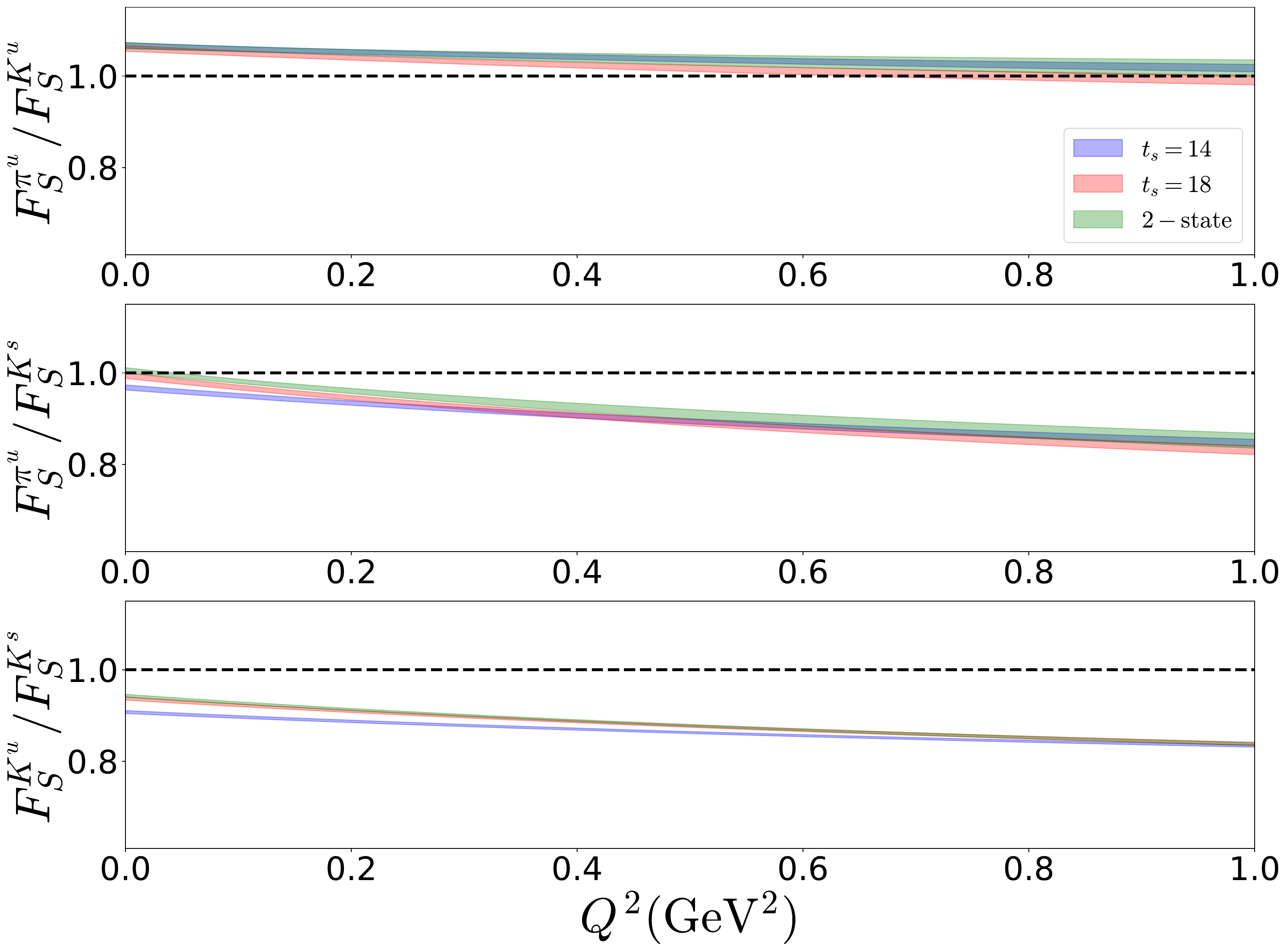}
    \caption{The ratio $F_S^{\pi^u}/F_S^{K^u}$ (top), $F_S^{\pi^u}/F_S^{K^s}$ (center), and $F_S^{K^u}/F_S^{K^s}$ (bottom) for the scalar form factor as a function of $Q^2$ using the results obtained from both frames. The results for $t_s/a=14,\,18$ and the two-state fit are shown with blue, red and green bands, respectively.}
    \label{fig:SU3_rest_scalar}
\end{figure}

\begin{figure}[h!]
    \centering
    \includegraphics[scale=0.24]{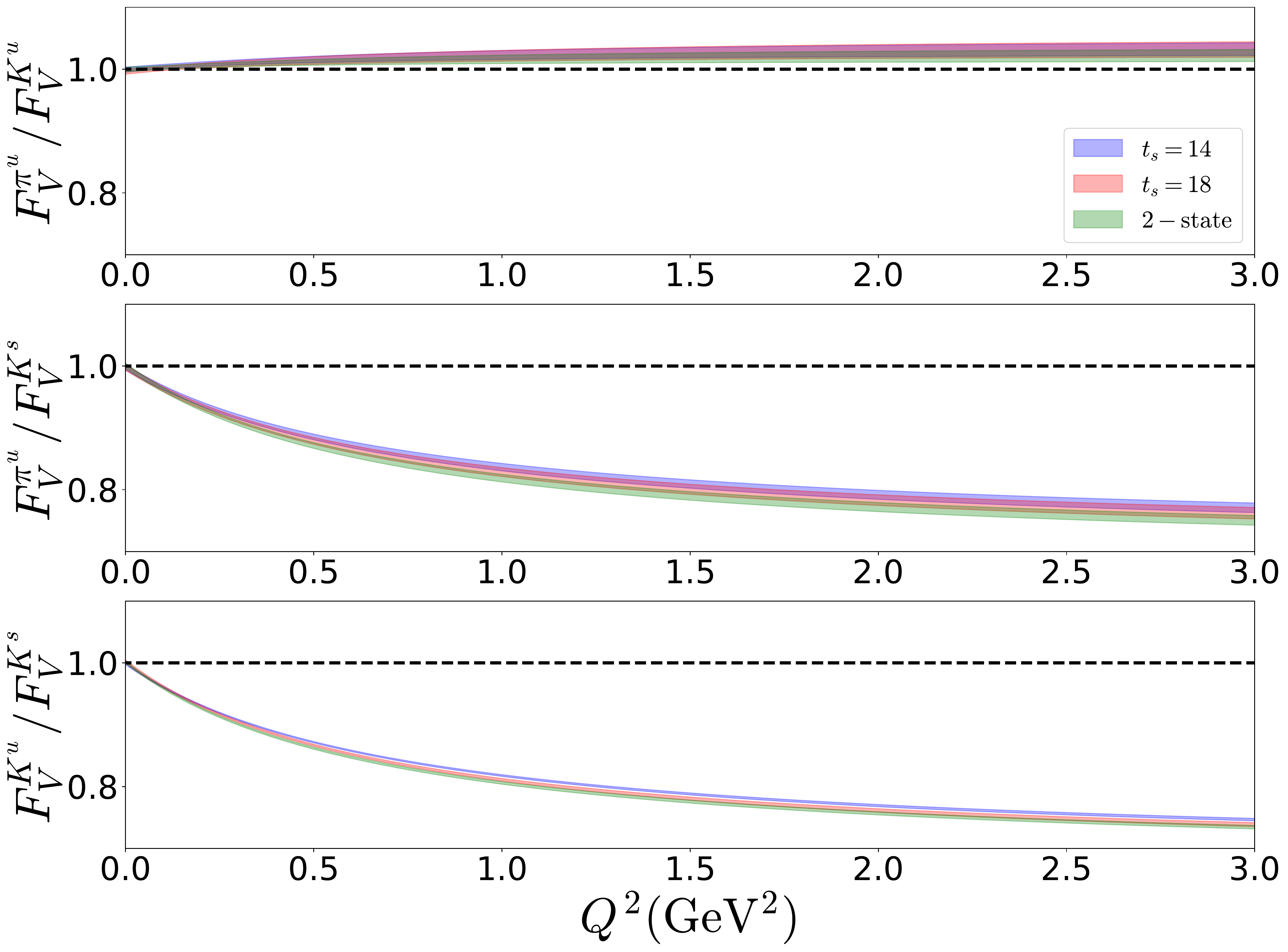}
    \caption{The ratio $F_V^{\pi^u}/F_V^{K^u}$ (top), $F_V^{\pi^u}/F_V^{K^s}$ (center), and $F_V^{K^u}/F_V^{K^s}$ (bottom) for the vector form factor as a function of $Q^2$ using the results obtained from both frames. The results for $t_s/a=14,\,18$ and the two-state fit are shown with blue, red and green bands, respectively.}
    \label{fig:SU3_vector}
\end{figure}

\begin{figure}[h!]
    \centering
    \includegraphics[scale=0.24]{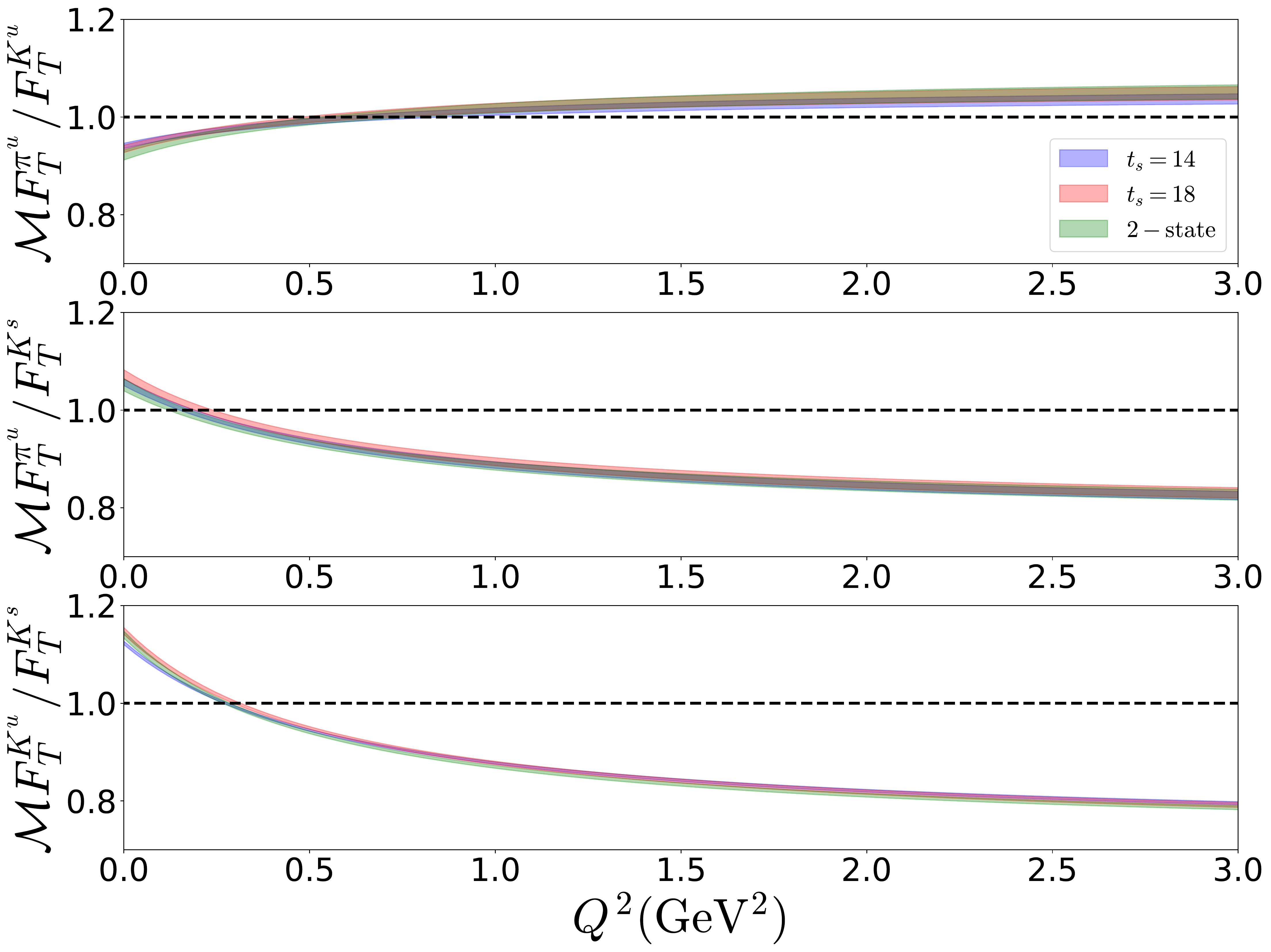}
    \caption{The ratio $F_T^{\pi^u}/F_T^{K^u}$ (top), $F_T^{\pi^u}/F_T^{K^s}$ (center), and $F_T^{K^u}/F_T^{K^s}$ (bottom) for the tensor form factor as a function of $Q^2$ using the results obtained from both frames. The results for $t_s/a=14,\,18$ and the two-state fit are shown with blue, red and green bands, respectively.}
    \label{fig:SU3_tensor}
\end{figure}

Studying both the pion and kaon form factors is useful to understand SU(3) flavor symmetry breaking effects. Such effects have been observed in nature, for example, in the charge radii of $\pi^{\pm}$ and $K^\pm$, as well as in $\pi^{0}$ and $K^0$. We have previously investigated such effects in the Mellin moments of the pion and kaon PDFs and their reconstructed PDFs~\cite{Alexandrou:2021mmi}. In this work, we draw conclusions on the SU(3) flavor symmetry breaking using the scalar, vector and tensor form factors. To this end, we examine the ratios $F^{\pi^u}/F^{K^u}$, $F^{\pi^u}/F^{K^s}$, and $F^{K^u}/F^{K^s}$ for each form factor. Note that the numerical value of $Q^2$ depends on the mass of the meson as given in Eqs.~(\ref{eq:Q2_r}) - (\ref{eq:Q2_b}). Thus, the pion and kaon form factors are extracted at different values of $Q^2$. Therefore, to study the SU(3) flavor symmetry breaking, we use the fitted values of the form factors, as detailed in Section~\ref{sec:fits_pion} for the pion and Section~\ref{sec:fits_kaon} for the kaon. We are also interested on how excited-states contamination affect such ratios, so we study them using the parametrizations on the individual plateau values, as well as the two-state fit. In Fig.~\ref{fig:SU3_rest_scalar} - \ref{fig:SU3_tensor} we show the aforementioned ratios for the scalar, vector and tensor operators. For a direct comparison, we keep the same y-axis for  $F^{\pi^u}/F^{K^u}$, $F^{\pi^u}/F^{K^s}$, and $F^{K^u}/F^{K^s}$ for a given form factor. To increase readability, we only show the results for $t_s/a=14,\,18$ and the two-state fit. For the case of the tensor form factor we include the ratio of the meson masses, ${\cal M}\equiv m_K/m_\pi$. For all cases, we use the parametrizations obtained from all lattice data in both the rest and boosted frames.
  
One of the common aspects for all the ratios is that there are no excited states contamination. Another common characteristic is that the ratios $F_{S, V, T}^{\pi^u}/F_{S, V, T}^{K^u}$ have very mild dependence on $Q^2$ and are close to unity. Regardless of the form factor, the up-quark contribution in the kaon becomes about 80$\%$ of that of strange quark as $Q^2$ increases. Since the up-quark component is similar in the pion and kaon, we also find that the up-quark contribution in the pion is approximately  80$\%$ that of the strange quark in the kaon.

\section{Transverse spin of the pion and kaon}
\label{sec:impact_b}

The computation of both the vector and tensor form factors can be used as a probe of the transverse spin structure of hadrons and the Boer-Mulders effect in the pion and kaon. Ref.~\cite{Burkardt:2000za} discusses how GPDs at zero skewness contain physical interpretation when Fourier transformed into the impact-parameter space. In this representation, they describe the spatial distributions in the transverse plane of partons with fixed longitudinal momentum. This allows one to extract the densities of transversely polarized quarks in the hadron under study~\cite{Diehl:2005jf}. It is worth mentioning that the Sivers asymmetry arises from the asymmetry of the quark distribution in the impact parameter space. The above observations can also be extended for the moments of GPDs, that is the form factors and their generalizations. Here, we outline the procedure for the lowest moments, and use our lattice data for a numerical implementation. 

The density for the lowest moment is given by
\begin{equation}
\label{eq:density}
\rho(b_\perp,s_\perp) = \frac{1}{2} \Big[ F_V(b_\perp^2) - \frac{s_\perp^i \epsilon^{ij} b_\perp^j}{m}\, \frac{\partial F_T(b_\perp^2)}{\partial b_\perp^2}  \Big]\,,    
\end{equation}
where $s_\perp$ is the quark transverse spin vector, and $b_\perp= (b_x,b_y) =  (b\, \cos{\phi}, b\,\sin{\phi})$.
$F_{\cal O}(b_\perp^2)$ is a Fourier transform of the form factor $F_{\cal O}(t)$ with momentum transfer squared in the transverse direction,
\begin{equation}
\label{eq:FT}
 F_{\cal O}(b_\perp^2) = \frac{1}{2\pi} \int d^2 \Delta_\perp^2 e^{-i b_\perp \cdot \Delta_\perp} \,F_{\cal O}(t=-\Delta_\perp^2)\,.
\end{equation}
One can take advantage of the parametrization of Eq.~\eqref{eq:fit} to extract a continuous function of $Q^2$ for the form factors. For the monopole Ansatz, the impact parameter form factor and its derivative become
\begin{eqnarray}
F_{\cal O}(b_\perp^2) = 
 \frac{M_{\cal O}^2 \, F_{\cal O}(0)}{2\pi} K_0(M_{\cal O} \,\sqrt{b_\perp^2}) \,, \qquad
\frac{\partial F_{\cal O}(b_\perp^2)}{\partial b_\perp^2} = -
 \frac{M_{\cal O}^3 \, F_{\cal O}(0)}{4\pi \sqrt{b_\perp^2}} K_{-1}(M_{\cal O} \,\sqrt{b_\perp^2})\,,
\end{eqnarray}
where $K_n(x)$ are the modified Bessel functions with $K_{-1}(x)=K_1(x)$. It should be noted that the parametrization of the form factors alleviates the issue of applying a Fourier transform on discretized lattice data. However, there are constraints related to the maximum $Q^2$ value that the form factors can be obtained. Indeed, in our calculation, the use of the rest frame gives access in the pion (kaon) form factors up to $\sim 0.5$ GeV$^2$ ($\sim 1$ GeV$^2$). The use of the boosted frame allows to better parametrize the $Q^2$ dependence via an extended data set up to $\sim 2.5$ GeV$^2$ for the pion and $\sim 3$ GeV$^2$ for the kaon. Therefore, the resolution of the impact parameter~\cite{Diehl:2002he}, $1/\sqrt{Q^2_{\rm max}}$, is of the order of 0.1 fm in this work.

Combining all the above, the density can be written in terms of the fit parameters obtained in sections ~\ref{sec:fits_pion} for the pion and \ref{sec:fits_kaon} for the kaon,
\begin{equation}
\label{eq:density2}
\rho(b_\perp,s_\perp) = \frac{1}{2} \Big[
\frac{M_V^2 \, F_V(0)}{2\pi} K_0(M_V \,\sqrt{b_\perp^2}) 
+ \frac{s_\perp^i \epsilon^{ij} b_\perp^j}{m}\, 
\frac{M_T^3 \, F_T(0)}{4\pi \sqrt{b_\perp^2}} K_{-1}(M_T\, \sqrt{b_\perp^2}) \Big]\,.  
\end{equation}
For quarks polarized in the x or the y direction, Eq.~\eqref{eq:density2} can be used to find the average transverse shift in the other perpendicular direction. For instance, if the polarization is along the x axis, $s_\perp=(1,0)$, then the average transverse shift in the y direction is defined as
\begin{equation}
\langle b^y_\perp\rangle = \frac{\int d^2 b_\perp b_\perp^y \rho(b_\perp,s_\perp)}{\int d^2 b_\perp \rho(b_\perp,s_\perp)}    \,.
\end{equation}
Using the $Q^2$-parametrization of the form factors, the above simplifies to
\begin{equation}
\langle b^y_\perp\rangle = \frac{F_T(0)}{2 m F_V(0)}\,.
\end{equation}
Here, we use the actual data for the vector form factor at $Q^2=0$. For $F_T(0)$ we use the value obtained from the monopole fit using the lattice data from both frames and $Q^2$ up to 2.5 and 3 GeV$^2$ for the pion and kaon, respectively. Furthermore, we use the two-state fit results to suppress any excited-states contributions. For the pion we find
\begin{equation}
\label{eq:pion_by}
\langle b^y_\perp\rangle^u_\pi = 0.1373(17) \,{\rm fm}\,,
\end{equation}
and for the kaon
\begin{equation}
\langle b^y_\perp\rangle^u_K = 0.1465(9) \,{\rm fm}\,,\qquad
\langle b^y_\perp\rangle^s_K = 0.1287(6) \,{\rm fm}\,.
\label{eq:kaon_by}
\end{equation}

\begin{figure}[h!]
    \centering
    \includegraphics[scale=0.59]{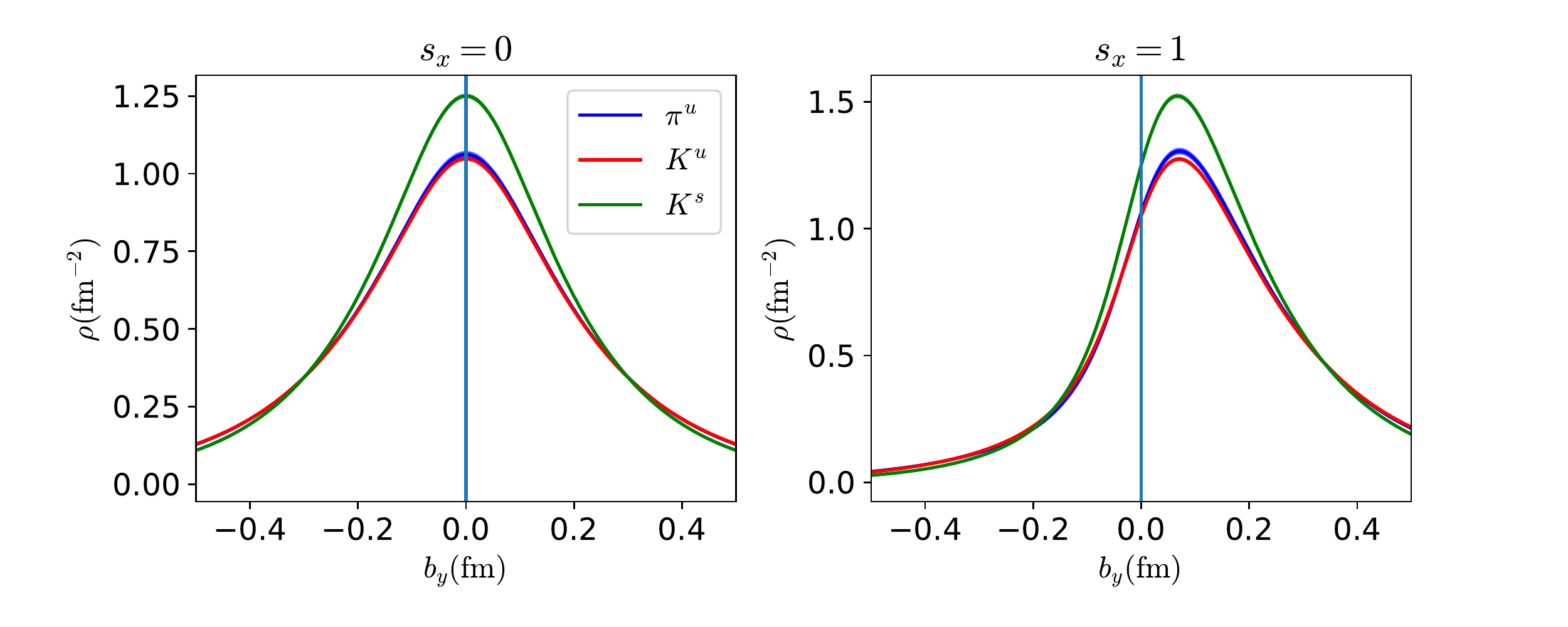}
    \vspace*{-0.4cm}
    \caption{Profile of the density $\rho(b_\perp,s_\perp)$ for unpolarized quarks, $s_\perp=(0,0)$, (left panel) and for transversely polarized quarks in the x direction, $s_\perp=(1,0)$,  (right panel) in the pion and kaon. $b_x$ was set to 0.15 fm. Red, blue, green bands correspond to up quark in the pion, the up quark in the kaon, and the strange quark in the kaon, respectively. }
        \label{fig:densities3}
\end{figure}

In Fig.~\ref{fig:densities3} we show the profile of the density plot for unpolarized quarks ($s_\perp=(0,0)$) and for transversely polarized quarks in the x direction ($s_\perp=(1,0)$) in the pion and kaon. For presentation purposes we set $b_x=0.15$ and plot the density as a function of $b_y$. In the  unpolarized case, it is symmetric with respect to $b_x$ and $b_y$, as can be seen from the first term of Eq.~\eqref{eq:density2}. On the contrary, in all cases of polarized quarks we find an asymmetry in $b_y$, with the maximum of the density at about $b_y=0.07$ fm. The densities for the up quarks are almost the same in the pion and the kaon. The density of unpolarized strange quarks is larger than for unpolarized up quarks in the region $|b_y|<0.3$ fm. A similar picture is observed for the polarized quarks with the region of dominance of the strange quarks is shifted in $-0.15 {\rm \,fm} < b_y <0.35$ fm.
 \vspace*{-0.3cm}

\begin{figure}[h!]
    \centering
    $s_x=0$\hspace*{5cm} $s_x=1$\\
    \includegraphics[scale=0.43]{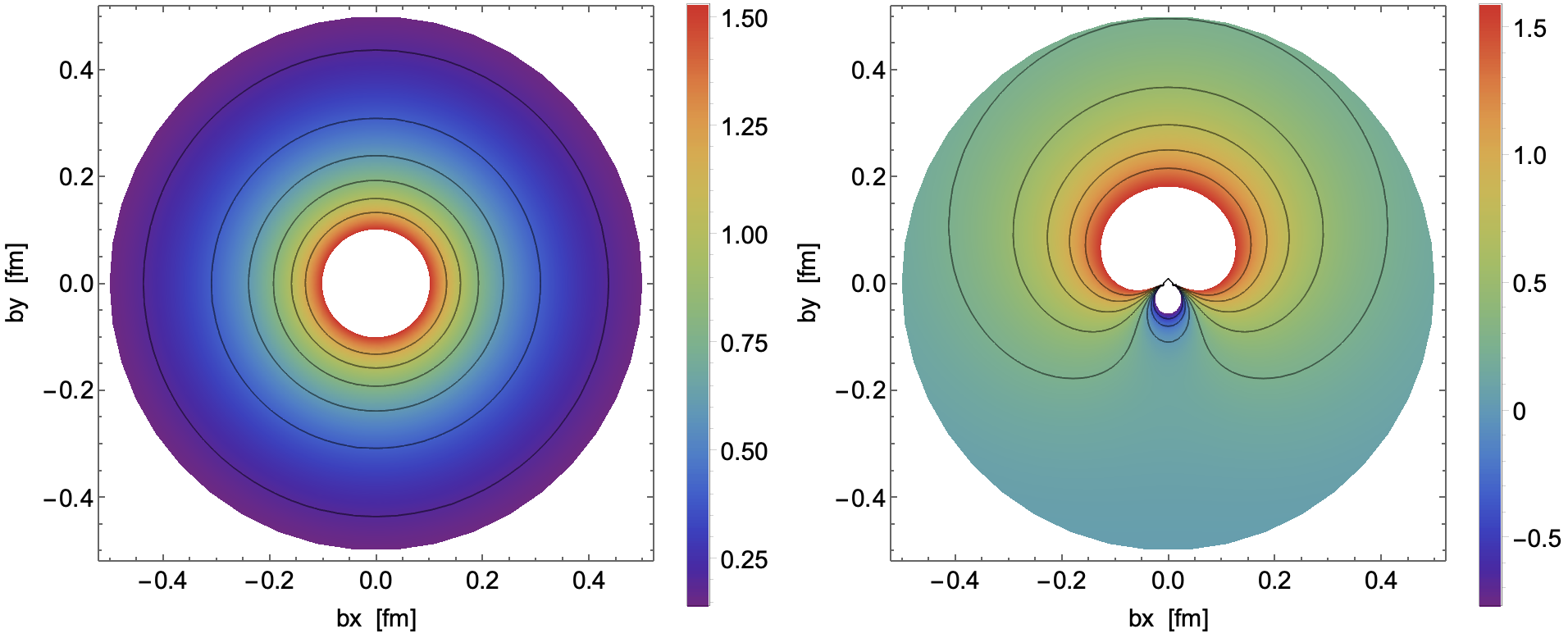}\\
    \includegraphics[scale=0.43]{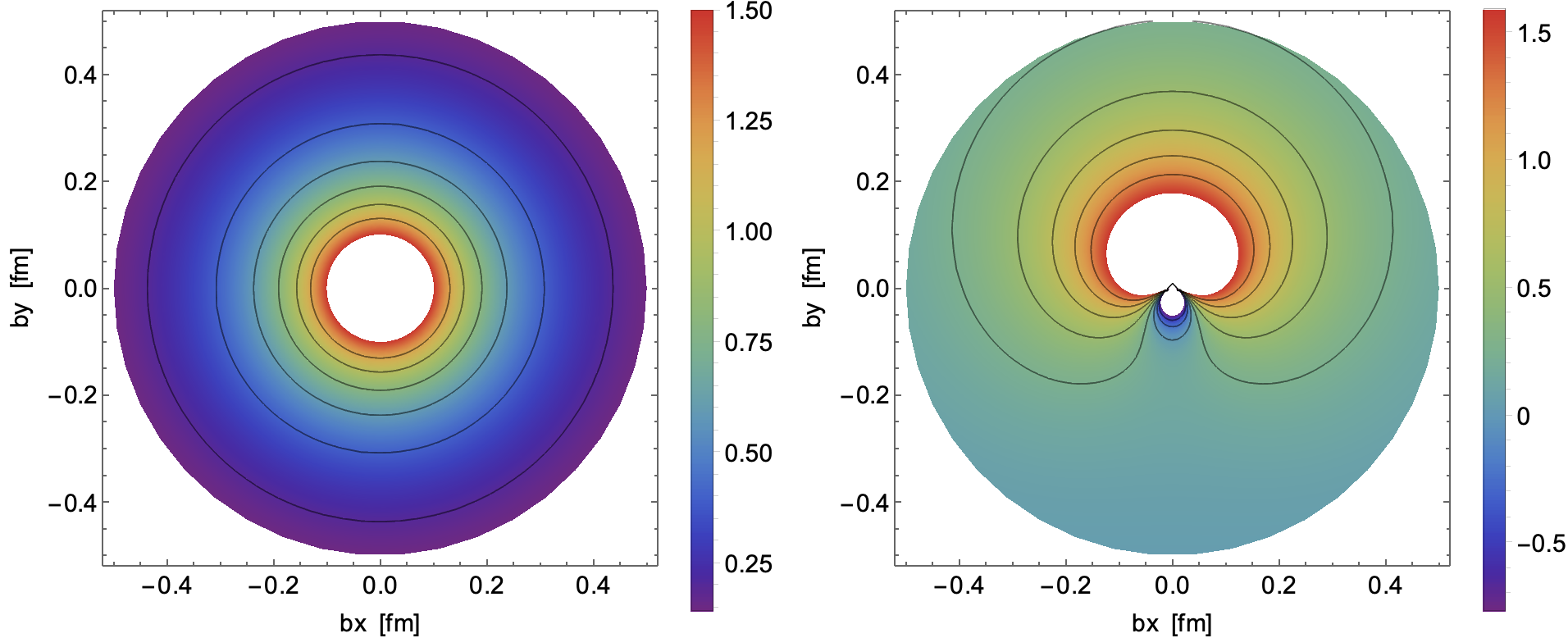}\\
    \includegraphics[scale=0.43]{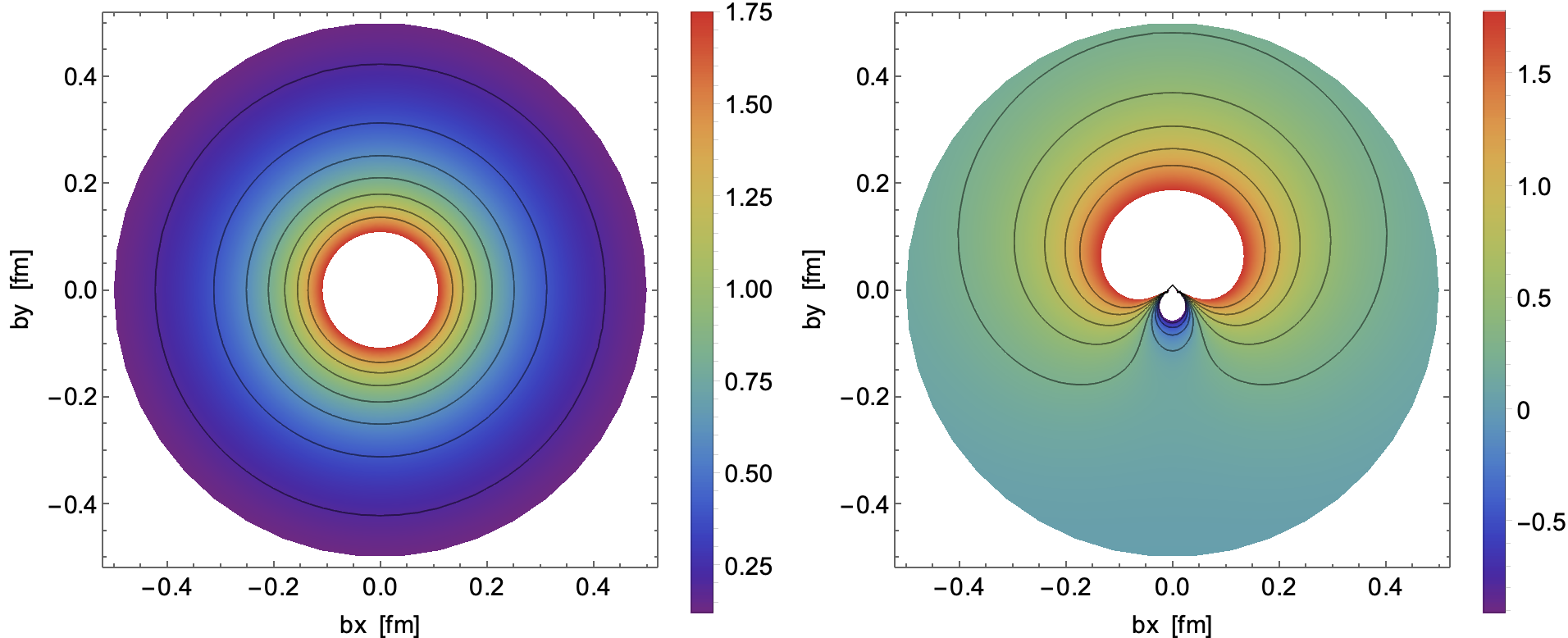}
    \caption{Density plot of Eq.~\eqref{eq:density2} of unpolarized quark (left) and transversely polarized quarks in the x direction (right). From top to bottom we show the case of the up quark in the pion, the up quark in the kaon, and the strange quark in the kaon. The white area in the center represents a value of the density outside the regions given with red and purple color.}
    \label{fig:densities1}
\end{figure}

The effect of distortion for polarized quarks discussed above can also be seen in Fig.~\ref{fig:densities1}, in which we plot the density using the parametrization $b_\perp=|b_\perp|(\cos{\phi},\sin{\phi})$. For the form factors at $Q^2=0$ and the monopole masses we use the mean values given in Eqs.~\eqref{eq:pion_fit_V} - \eqref{eq:pion_fit_T} for the pion, Eqs.~\eqref{eq:kaon_fit_V_u} - \eqref{eq:kaon_fit_T_u} for the up-quark in the kaon, and Eqs.~\eqref{eq:kaon_fit_V_s} - \eqref{eq:kaon_fit_T_s} for the strange-quark in the kaon. The picture for up and strange quarks with the same polarization is qualitatively the same.

\begin{figure}[h!]
    \centering
    $s_x=0$\hspace*{8cm} $s_x=1$ \hspace*{2cm}\\
    \includegraphics[scale=0.5]{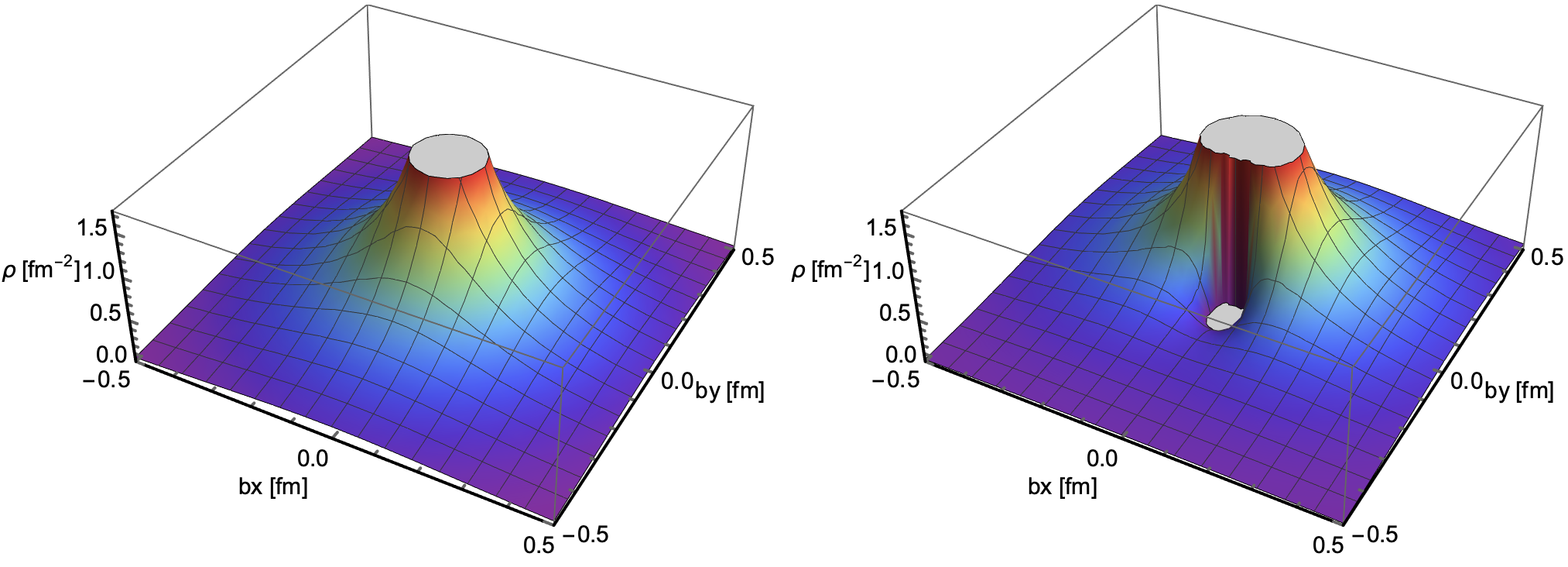}\\
    \includegraphics[scale=0.5]{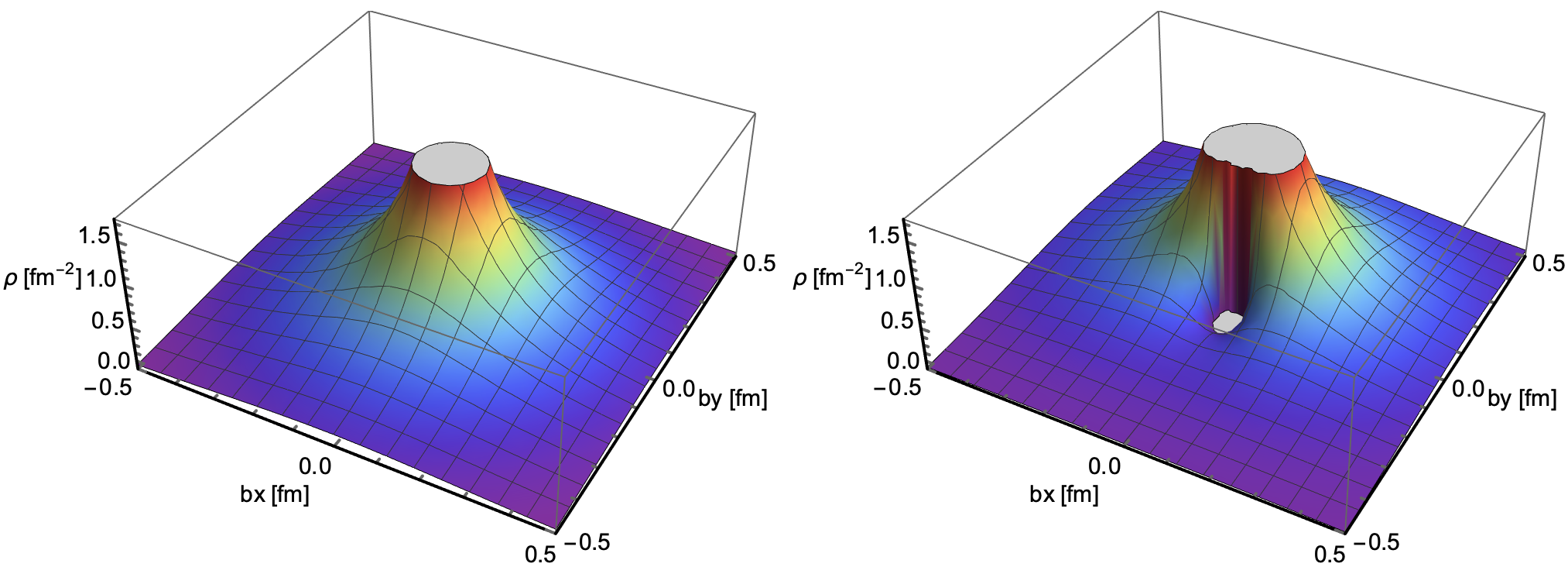}\\
    \includegraphics[scale=0.5]{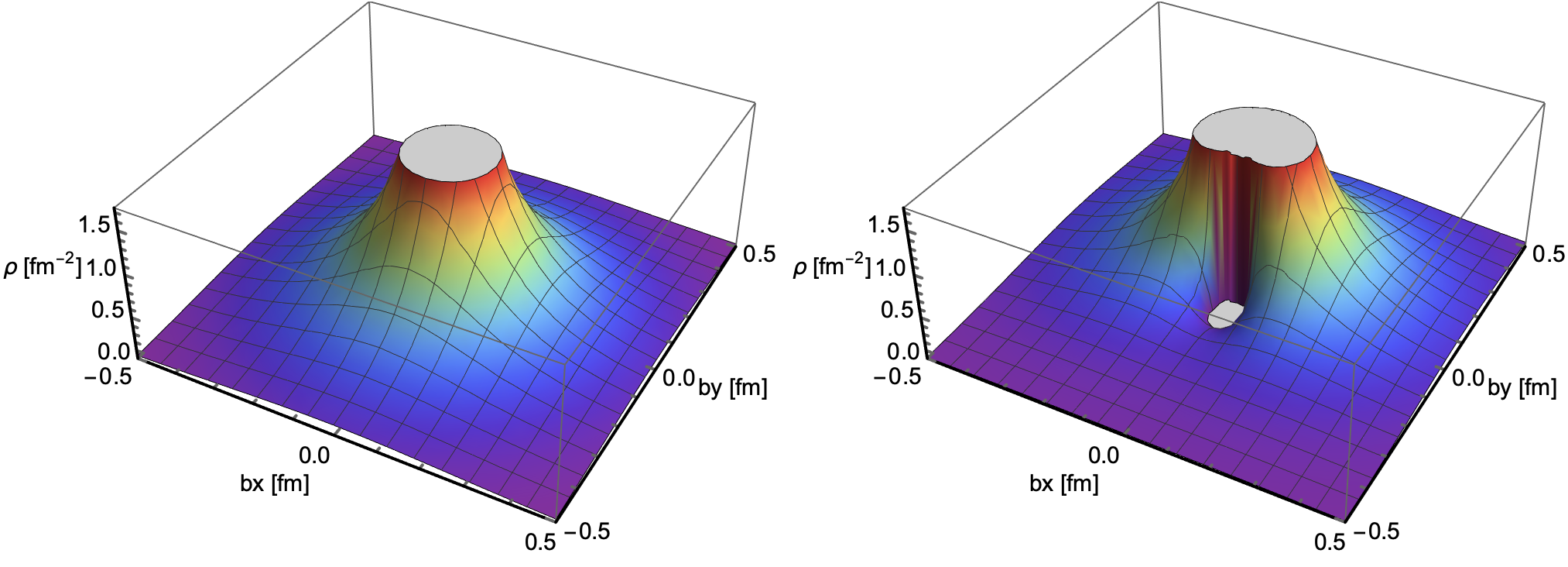}
    \caption{Eq.~\eqref{eq:density2} of unpolarized quark (left) and transversely polarized quarks in the x direction (right) as a function of $b_y$ for $b_x=0.15$. From top to bottom we show the up-quark for the pion, up-quark for the kaon, and strange-quark for the kaon. The gray area represents an increase in the density outside the values indicated by the range of the vertical axis.}
    \label{fig:densities2}
\end{figure}

Similar conclusions are drawn from the three-dimensional plots of the density as a function of $b_x$ and $b_y$, shown in Fig.~\ref{fig:densities2} using the same data as in Fig.~\ref{fig:densities1}. The overall picture for the pion is qualitatively compatible with the work of Ref.~\cite{Brommel:2007xd}, which employs several ensembles of $N_f = 2$ improved Wilson fermions with pion mass ranging between 400 MeV and 1 GeV.

\section{Summary and conclusions }
\label{sec:summary}

We present a calculation in lattice QCD of the scalar, vector and tensor form factors of the pion and kaon. Our results are obtained from the analysis of  an $N_f=2+1+1$ ensemble of twisted mass fermions with clover improvement  with 260 MeV pion mass and 530 MeV kaon mass. The scalar and tensor form factors are renormalized non-perturbatively and the results are given in the $\overline{\rm MS}$ scheme at a scale of 2 GeV. For the vector form factor we use the conserved vector operator, which does not need renormalization. 

We obtain the form factors on two kinematic setups, that is, the rest frame and a momentum-boosted frame of 0.72 GeV ($\mathbf{p'}=\frac{2\pi}{L}(\pm1,\pm1,\pm1)$). The use of the boosted frame has the advantage of a wider and denser range of values for the four-vector momentum transfer squared, $-t=Q^2$. This becomes possible due to the higher ground-state energy compared to the rest frame. However, much larger statistics are needed to control statistical uncertainties. Here we use a factor of 50 more statistics in the boosted frame compared to the rest frame. This allows the extraction of the pion form factors up to $Q^2=2.5$ GeV$^2$ and the kaon form factors up to $Q^2=3$ GeV$^2$. The form factors are frame independent, so one can combine the data obtained from both the rest and boosted frame. Overall, we find that there is excellent agreement between the two frames for the vector form factors of the pion and kaon, as well as the strange-quark contributions for the scalar and tensor form factors of the kaon. The aforementioned agreement mostly holds for the two-state fits, as the excited-states contamination affects differently the data in each frame. For the up-quark part of the pion and kaon scalar and tensor form factors we find agreement in the small-$Q^2$ region, with some deviations in the slope between 0.25 - 0.5 GeV$^2$ for the pion and 0.35 - 1 GeV$^2$ for the kaon. This is an indication of  cutoff effects that would need ensembles at least three lattice spacings to quantify.

In this work we limit ourselves to investigating possible sources of systematic uncertainties that can be studied on a single ensemble, the primary one being excited-states contamination. To this end, we produce data for six values of the source-sink time separation in the rest frame ($1.12-2.23$~fm) and four values in the boosted frame ($1.12-1.67$~fm). The two- and three-point functions are analyzed using single-state and two-state ansatz. We give our final results for the form factors using the two-state fits, and we parametrize their $Q^2$ dependence using a monopole fit. This leads to the scalar, vector and tensor monopole masses, and their corresponding radii. For the tensor form factor we also extract the tensor anomalous magnetic moment, $k_T$ which can only be obtained from fits on the data. Another systematic effect that we study is the sensitivity of the extracted parameters on the fit range of $Q^2$ and the frame sets included in the fit. As expected from the comparison of the form factors in the two frames, there are some tensions in extracting the scalar and tensor monopole masses and radii based on the data sets included in the fit. Our final results for the monopole masses and the tensor magnetic moment use all available lattice data in both frames. For the radii, we use all data up to $Q^2=0.5$ GeV$^2$ for the pion and $Q^2=1$ GeV$^2$ for the kaon. In all cases, we assign a systematic error by varying the fit range. The final results can be found in Eq.~\eqref{eq:pion_fit_S} - \eqref{eq:pion_fit_rT} for the pion and Eq.~\eqref{eq:kaon_fit_S} - \eqref{eq:kaon_fit_rT} for the kaon.

We compare the parametrized form factors for the pion and kaon to address SU(3) flavor symmetry breaking effects. Our analysis indicates that excited states are suppressed in such ratios. We also find a mild $Q^2$-dependence in the case of $F^{\pi^u}/F^{K^u}$ for all operators with the ratios being around the value 1. For the $F^{\pi^u}/F^{K^s}$ and $F^{K^u}/F^{K^s}$ cases, we find sizeable $Q^2$-dependence and SU(3) flavor symmetry breaking effects up to 20$\%$. Finally, combining the data for the vector and tensor form factors we also obtain the lowest moment of the densities of unpolarized and transversely polarized quarks in the impact parameter space, $b_\perp$. As expected, a distortion appears for the polarized case with the density reaching maximum for positive values of $b_y$. The values for the average transverse shift in the y direction for polarized quarks in the x direction are given in Eqs.~\eqref{eq:pion_by} - \eqref{eq:kaon_by}.

An extension of this work, will be the investigation of other sources of systematic uncertainties, such as volume and discretization effects. In the near future, we will perform such calculations on ensembles with a physical pion mass. We also plan to advance this work, as well as the work of Ref.~\cite{Alexandrou:2020gxs} with the calculation of the generalized form factors of the one-derivative operators using a similar setup as for the form factors.

\vspace*{0.25cm}

 \begin{acknowledgements}

We would like to thank all members of ETMC for a very constructive and enjoyable collaboration. M.C. thanks Martin Hoferichter for interesting discussions on Ref.~\cite{Hoferichter:2018zwu}.  
M.C. and J.D. acknowledge financial support by the U.S. Department of Energy Early Career Award under Grant No.\ DE-SC0020405. 
K.H. is financially supported by the Cyprus Research Promotion foundation under contract number POST-DOC/0718/0100, CULTURE/AWARD-YR/0220 and EuroCC project funded by the Deputy Ministry of Research, Innovation and Digital Policy, the Cyprus The Research and Innovation Foundation and
the European High-Performance Computing Joint Undertaking (JU) under grant agreement No. 951732. The JU received
support from the European Union’s Horizon 2020 research and innovation programme.
S.B. is supported by the H2020 project PRACE 6-IP (grant agreement No 82376) and the EuroCC project (grant agreement No. 951732).
C.L. is supported by the Argonne National Laboratory with a research subcontract with Temple University.
A.V. is supported by the U.S. National Science Foundation under Grants No. PHY17-19626 and PHY20-13064.
This work was in part supported by the U.S. Department of Energy, Office of Science, Office of Nuclear Physics, contract no.~DE-AC02-06CH11357 and the European Joint Doctorate program STIMULATE funded from the European Union’s Horizon 2020 research and innovation programme under grant agreement No 765048.
This work used computational resources from Extreme Science and Engineering Discovery Environment (XSEDE), which is supported by National Science Foundation grant number TG-PHY170022. 
It also includes calculations carried out on the HPC resources of Temple University, supported in part by the National Science Foundation through major research instrumentation grant number 1625061 and by the US Army Research Laboratory under contract number W911NF-16-2-0189. 
\end{acknowledgements}

\bibliography{references.bib}

\end{document}